\documentclass{scrartcl}

\usepackage{amssymb}
\usepackage{amsmath}
\usepackage{graphicx}
\usepackage{xcolor}
\usepackage[round,authoryear]{natbib}
\usepackage[colorlinks,linkcolor=black,anchorcolor=black,citecolor=black,urlcolor=blue]{hyperref}
\usepackage{booktabs} 
\usepackage{multirow} 
\usepackage{url}
\usepackage{rotating}

\setcitestyle{comma,aysep={}}

\title{On scenario construction for stochastic shortest path problems in real road networks}

\author{Dongqing Zhang\thanks{Business School, Sichuan University, Chengdu 610065, China}, Stein W. Wallace\thanks{NHH Norwegian School of Economics, Bergen, Norway}, Zhaoxia Guo\thanks{Corresponding author; Email: zx.guo@alumni.polyu.edu.hk; Business School, Sichuan University, Chengdu 610065, China}, \\Yucheng Dong\thanks{Corresponding author; Email: ycdong@scu.edu.cn; Business School, Sichuan University, Chengdu 610065, China}, Michal Kaut\thanks{SINTEF Technology and Society, Trondheim, Norway}}

\date{\today}

\begin{document}

\maketitle

\begin{abstract}
Stochastic shortest path computations are often performed under very strict time constraints, so computational efficiency is critical. A major determinant for the CPU time is the number of scenarios used. We demonstrate that by carefully picking the right scenario generation method for finding scenarios, the quality of the computations can be improved substantially over random sampling for a given number of scenarios. We study a real case from a California freeway network with 438 road links and 24 5-minute time periods, implying 10,512 random speed variables, correlated in time and space, leading to a total of 55,245,816 distinct correlations. We find that (1) the scenario generation method generates unbiased scenarios and strongly outperforms random sampling in terms of stability (i.e., relative difference and variance) whichever origin-destination pair and objective function is used; (2) to achieve a certain accuracy, the number of scenarios required for scenario generation is much lower than that for random sampling, typically about 6-10 times lower for a stability level of 1\%; and (3) different origin-destination pairs and different objective functions could require different numbers of scenarios to achieve a specified stability.


\end{abstract}

\section{Introduction}

The shortest path problem, with variants such as the quickest path problem, is a classical combinatorial optimization problem, which is to find a path in a network (graph) from an origin node to a destination node with shortest link length or quickest travel time. Due to its diverse applications like route planning in road networks, timetabling for railways, or scheduling for airplanes \citep{Delling2012}, the shortest path problem has been the subject of extensive research for many years \citep{Deo1984,Madkour2017}. Most papers deal with deterministic problems, see reviews \citep{Fu2006,Sommer2014,Bast2016} and relevant papers in recent years \citep{Zhang2017a,Hu2019,Brown2019}.

Real-world shortest path problems are stochastic in nature due to unpredictable factors such as traffic accidents, traffic control, and weather conditions. Hence, a large variety of stochastic shortest path (SSP) papers exist \citep{Chen2013,Wu2015,Cao2016,Rambha2016,Chen2017}. The majority of these assume that speeds on different road links and across different time periods are uncorrelated \citep{Murthy1996,Miller-Hooks2000,Nielsen2014,Prakash2018}. But in recent years, an increasing number of papers consider spatially correlated speeds \citep{Xing2013,Zhang2017,Prakash2018a}, that is, the speed on one road link is correlated with the speed on certain other links.

It is reasonable to assume that there exist strong spatial and temporal correlation among speeds in real road networks, largely due to traffic flow propagations over time and space, or an event, for example dynamic traffic management \citep{Koester2018}, that affects traffic capacities in a wide area \citep{Rachtan2013}. We demonstrate that for our data set this is indeed the case. Temporal correlation in speeds means that the speed in one time period is correlated with the speed in other time periods. The spatial and temporal correlations have been studied by some researchers \citep{Cheng2012,Rachtan2013,Ermagun2017}, and considered in several travel time- and route-related decision-making problems \citep{Min2011,Zou2014,Tulic2014}. The significance of spatial and temporal correlation of stochastic speeds in SSP problems has also been demonstrated by \cite{Huang2012}, \cite{Zockaie2013}, and \cite{Zockaie2016}. \cite{Huang2012} and \cite{Zockaie2013} found that spatial and temporal correlations could affect the optimal path and the impact was related to the levels of correlation. \cite{Zockaie2016} discovered that the optimal path travel time distribution in the spatially and temporally correlated case fell between the uncorrelated case and the spatially correlated case.

However, the SSP research with both spatially and temporally correlated stochastic speeds (travel times) is still in its infancy, although there are several studies reported in recent years \citep{Huang2012,Zockaie2013,Yang2014,Zockaie2016,Huang2018}. These studies adopted sampling-based methods to handle the spatial and temporal correlations of speeds. \cite{Zockaie2013} sampled 1,000 scenarios from a multivariate normal distribution using the Monte Carlo method. \cite{Yang2014} took 10 days of real travel times from the freeways of San Diego as scenarios. \cite{Zockaie2016} generated 86 scenarios using the traffic simulator, DYNASMART-P \citep{Mahmassani2009}, based on 86 days of historical data on network-wide demand levels, weather conditions, incidents, and routing strategies in downtown Chicago. \cite{Huang2012,Huang2018} sampled 50 scenarios (support points) for link travel times from an assumed truncated multivariate normal distribution, and \cite{Huang2018} demonstrated, using sensitivity analysis, that the number of scenarios had an impact on the solution quality. These previous studies simply pick an arbitrary number of scenarios and do not evaluate the effectiveness of the sampled scenarios. On one hand, it is well-known that sampling-based methods usually need a large number of samples to reflect the distribution of the random variables well. On the other hand, given historical data or marginal distributions and correlations of the random variables, it is still not clear how to generate appropriate scenarios needed for SSP problems with spatially and temporally correlated speeds, except for using random sampling. It is also not clear how to evaluate the quality of the generated scenarios for different SSP problems.

It is thus worthwhile to investigate how many scenarios are needed to establish stable, trustworthy results based on sampling, and then propose a much more efficient method based on explicit scenario generation for these SSP problems in real road networks. In a recent work, Guo, Wallace and Kaut \citeyearpar{Guo2018} used a copula-based scenario generation approach \citep{Kaut2014} to handle spatially and temporally correlated speeds in vehicle routing problems. For the investigated vehicle routing problem with the expected overtime minimization objective, they only needed 15 scenarios to achieve an objective-function evaluation stability of 1\% for a case with 142 nodes, 418 road links and 60 time periods, leading to over 25,000 correlated random speeds. The approach is based on a simplified map of Beijing, but the stochasticity of travel speeds is assumed.

Following Guo, Wallace and Kaut \citeyearpar{Guo2018} and \citet{Huang2018}, this paper investigates various SSP problems with spatially and temporally correlated speeds based on real travel speeds from the Caltrans Performance Measurement System (PeMS) at \url{http://pems.dot.ca.gov/}. We establish that there are statistically significant correlations in time and space in this data set. In addition, we point out that whatever the correlation structure, there is a much better way to find scenarios than by using random sampling. We demonstrate that whether the goal is to obtain maximal quality for a given number of scenarios (because there is a limited execution time available), or a certain quality for a minimal number of scenarios (because accuracy is important), the copula-based scenario generation method is substantially better. This paper contributes to the SSP literature by

\begin{enumerate}
\item investigating realistic stochasticity of spatially and temporally correlated travel speeds,
\item presenting an efficient method to generate scenarios for SSP problems in real road networks, and
\item investigating if and how different origin-destination (O-D) pairs and different objective functions affect the required number of scenarios to achieve a certain accuracy in the calculations.
\end{enumerate}

This paper is organized as follows. In Section~\ref{sec:stochasticity}, realistic stochasticity of spatially and temporally correlated travel speeds is studied based on a large data set of real travel speeds from California. The purpose of this section is to demonstrate that there are large and statistically significant correlations in our data set, thereby demonstrating that discussing correlations is indeed meaningful. Scenario generation and solution evaluations are discussed in Section~\ref{sec:SG}, while the test case and experimental setting are presented in Section~\ref{sec:case}. Then, Sections~\ref{sec:diffODs}--\ref{sec:diffOBJs} show the experimental results and analyses with different O-D pairs and different objective functions, respectively. In Section~\ref{sec:conclusions}, conclusions are made and future research directions are discussed.

\section{Data-Driven Stochasticity of Spatially and Temporally Correlated Travel Speeds}
\label{sec:stochasticity}
The purpose of this section is to analyse our data set from California in light of spatial and temporal correlations. We shall establish that there are large and statistically significant correlations among the link speeds in the data set, thereby confirming that handling correlations in SSP calculations is meaningful and potentially important. To analyse the stochasticity of travel speeds in a road network, it is necessary to collect a large number of real travel speeds on each link of the network in many time periods as basic data and then pre-process the data (e.g., data cleaning and imputation) for further use. With the development of intelligent transportation systems and information technology, we can collect the raw real-time data of vehicle travel speeds from a large number of auxiliary instruments, e.g., inductive-loop detectors, laser radars, GPSs, cellular networks, cameras, and microwave detectors \citep{Zhang2011}. This paper uses the ready-made data of vehicle travel speeds in PeMS, which are already detected and imputed using the methods in Chen \citeyearpar{Chen2003a}. The dataset is complete and has no missing data. The time period of travel speeds is 5 minutes.

\begin{figure}
	\centering
	\includegraphics[width=0.8\textwidth]{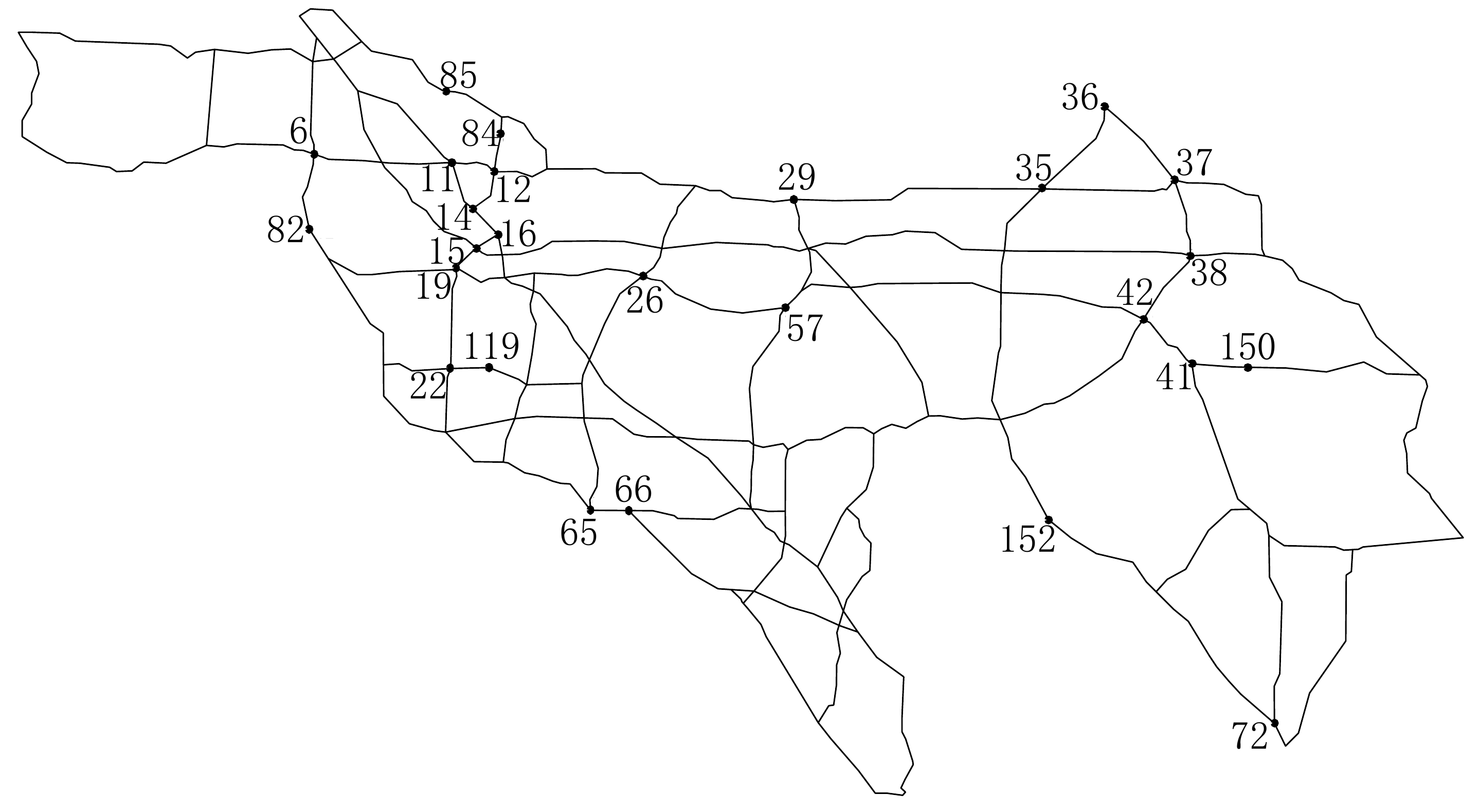}
\caption{Freeway network around Los Angeles. The numbers refer to nodes that we use in our O-D pairs.}
\label{fig:fig5}
\end{figure}

Since the real spatial and temporal correlations of stochastic travel speeds have rarely been reported in the SSP literature, we focus on this issue next. There exist very complicated spatial and temporal correlation between vehicle travel speeds in real road networks.
Take the Los Angeles spatio-temporal freeway network as an example.
Figure~\ref{fig:fig5} shows the freeway network, which consists of 173 nodes, 438 road links, and 3,417 speed detection stations. For presentation simplicity, we only present node numbers used in the main text in the figure. The road link distances were derived from a digital roadmap.
Then, to get the real freeway travel speed readings from the PeMS, we downloaded all speed readings of the 3,417 speed detection stations over 102 days, from May 1 to September 22 in 2017 excluding weekends and holidays.
In general, there is more than one detection station on each road link. Thus, the speed sample on one link in one time period is obtained by averaging the speed readings of all detection stations on this link in this time period.
The time period of downloaded travel speeds is 5 minutes, and the time range used throughout this paper is 2 hours, each therefore having 24 time periods. In each 2-hour time range, we have 10,512 ($438\times24$) random speed variables and 55,245,816 ($10,512 \times 10,511 \div 2$) distinct correlations.

\begin{figure}
\centering
\resizebox*{8cm}{!}{\includegraphics{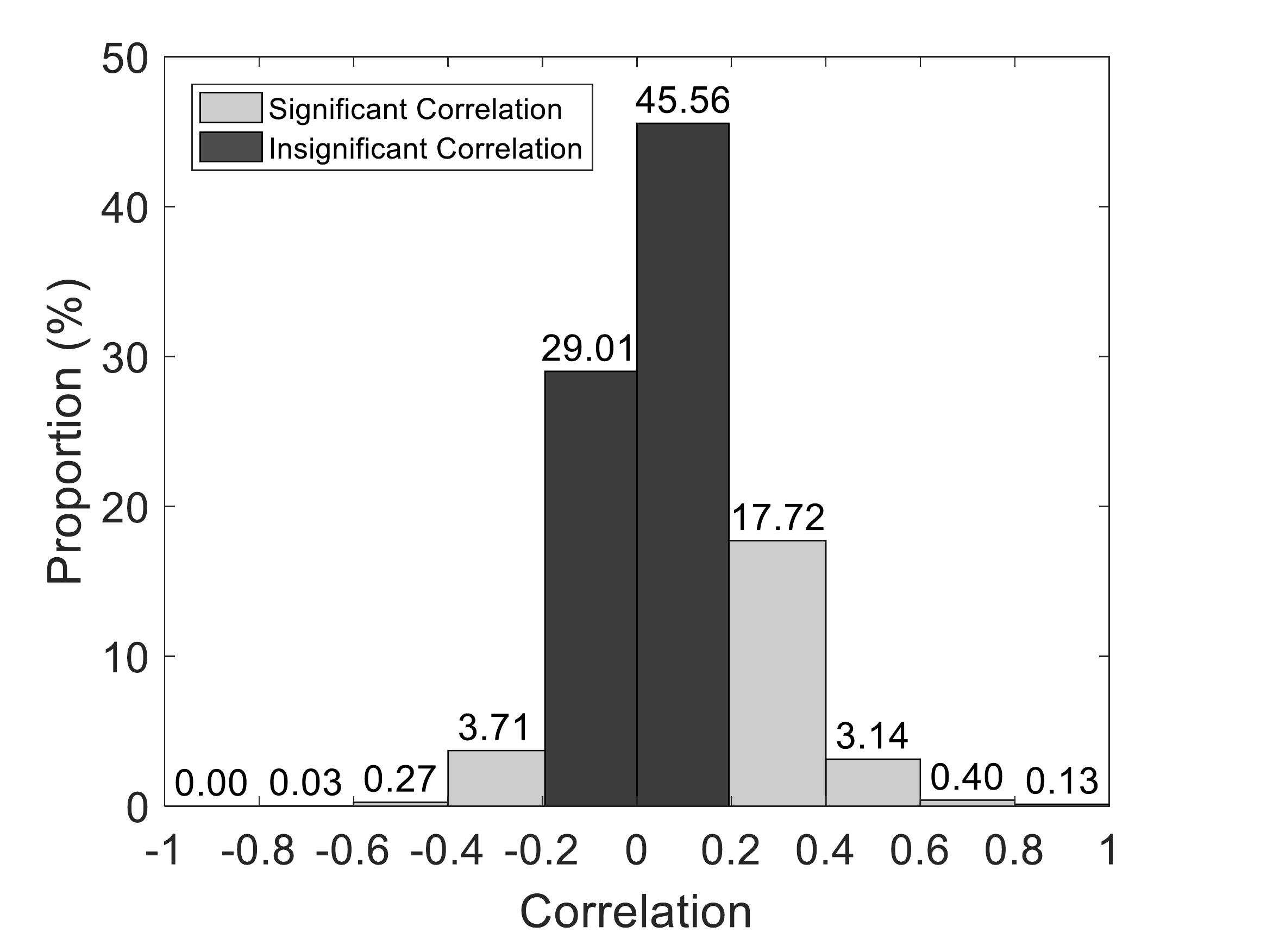}}
\caption{Frequency histogram of all correlations in Los Angeles freeway network between 8am and 10am.}
\label{fig:fig1}
\end{figure}

\begin{sidewaysfigure} 
\centering
\resizebox*{18cm}{!}{\includegraphics{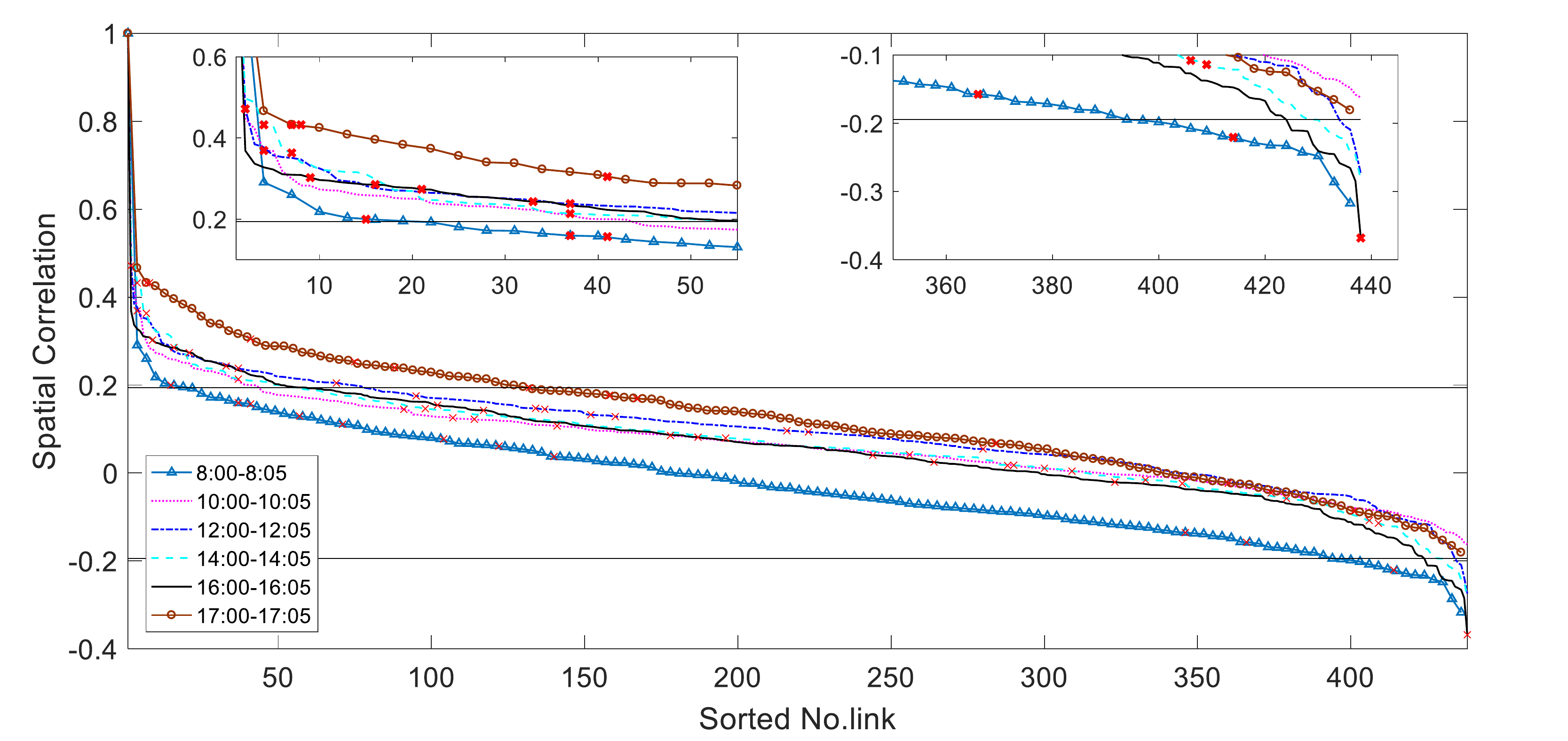}}
\caption{The spatial correlations of link (14, 11) with other links in the road network in six time periods. For the curve of each time period, the link numbers are sorted in the decreasing order of spatial correlations. The red crosses represent the correlations between link (14, 11) and its 11 neighbouring links (links sharing node 14 or 11). The correlations between the two horizontal lines are not significant with a significance level of 5\%, the rest are.}
\label{fig:fig2}
\end{sidewaysfigure}

\begin{figure}
\centering
\resizebox*{8cm}{!}{\includegraphics{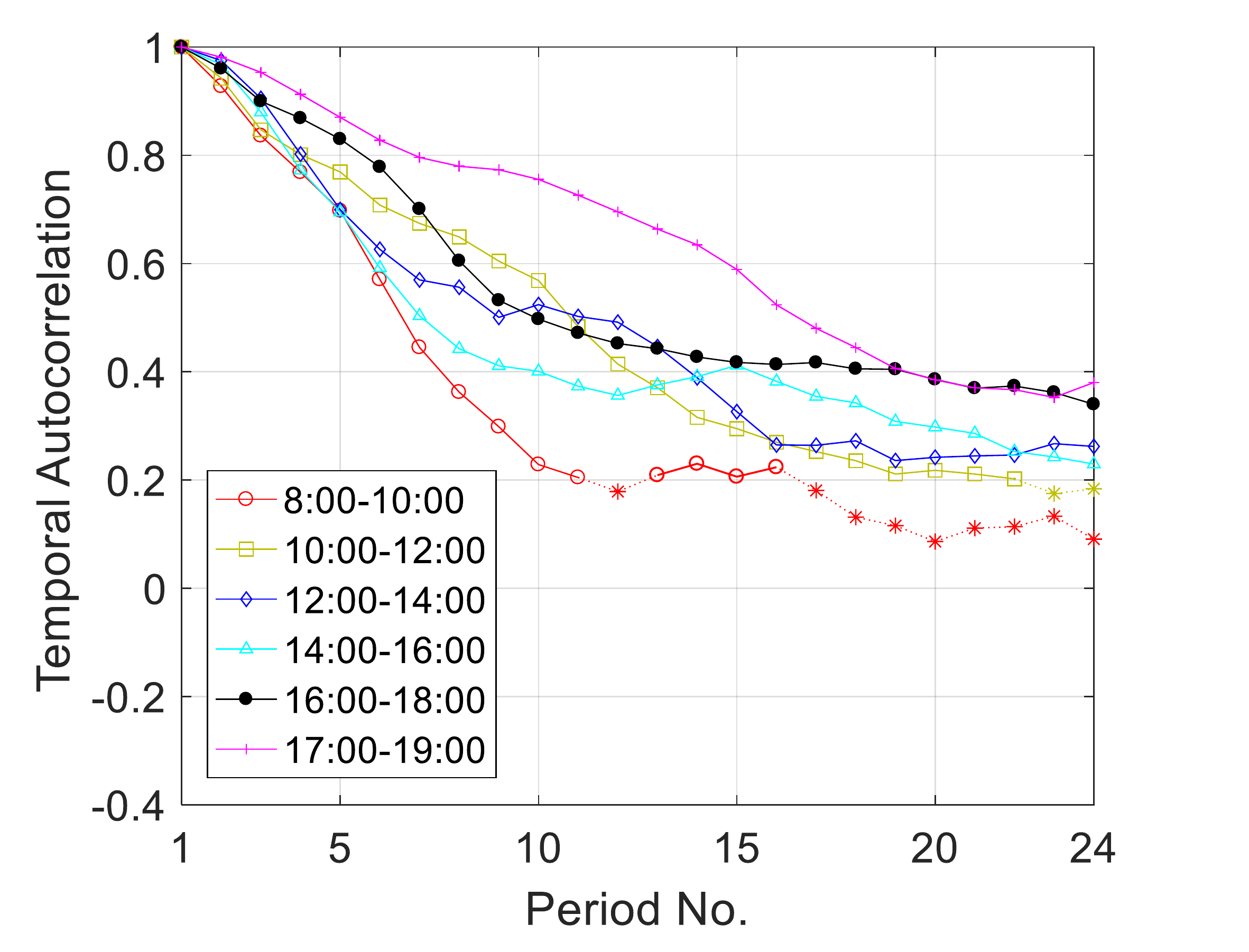}}
\caption{The temporal correlations of link (14, 11) in six time ranges. The stars in dashed lines are not significant at the 5\% level, the rest are.}
\label{fig:fig3}
\end{figure}

We have the following typical observations based on the real spatial and temporal correlations obtained.
\begin{enumerate}
  \item Figure~\ref{fig:fig1} shows the frequency histogram of all correlations between 8am and 10am. In this figure, the black bars represent the statistically insignificant correlations while other bars represent the statistically significant with a significance level of 0.05.
      According to the Student's t-test at the significance level of 5\%, we reject the null hypothesis that two variables are not correlated (i.e., $r$=0) at a confidence level of 95\%, if the $p$ value of t-statistic $t_r$ is less than or equal to 0.05. We thus have $t_r = |r-0|\times \sqrt{\frac{N-2}{1-r^2}} \geq 1.984$, where $N$ denotes the number of samples ($N$=102 in the test case). Therefore, two speeds are significantly correlated at a confidence level of 95\% if their correlation $r$ satisfies $|r| \geq 0.1946$. Otherwise, two speeds are insignificantly correlated (i.e., accept the null hypothesis). It can be seen from Figure~\ref{fig:fig1} that 74.57\% of the correlations are statistically insignificant. This figure also shows that only 0.56\% correlations have absolute value larger than 0.6. We have much more insignificant correlations because we consider all correlations among 24 time periods and all 438 links in the map. It is reasonable that links that are far from each other in time and space are not significantly correlated.

  \item Figure~\ref{fig:fig2} presents the spatial correlations between a randomly selected link (14, 11) and all other links in the network shown in Figure~\ref{fig:fig5} in six different time periods. In this figure, the correlations between the two horizontal lines that are parallel with the X-axis are statistically insignificant with a significance level of 0.05, and the red crosses represent the correlations between link (14, 11) and its 11 neighbouring links (links sharing node 14 or 11). We see that, in a given time period, the travel speeds on the link (14,11) could be either significantly or insignificantly correlated with those on its neighbouring links. This finding holds true on all 30 randomly chosen links we observed. We also found that (1) if two neighbouring links are in the same direction, the speeds on the links are usually significantly positively correlated; and (2) if two neighbouring links are in opposite directions, their speeds are often insignificantly correlated.
      Taking link (14, 11) in Figure~\ref{fig:fig5} as an example, the speeds on this link and its neighbour (12, 14) are negatively correlated (-0.22) in time period 8:00-8:05. The reason is simple. The two links and link (12, 11) form a triangle.
      A fair number of vehicles that are headed from node 12 to node 11, tend to choose the detour via node 14 when they observe that link (12,11) is congested and link (12,14) looks faster. However, these vehicles could slow down link (14,11) since many vehicles are travelling on this link. Contrariwise, rather few vehicles going from node 12 to node 11 choose the detour via node 14 if link (12,14) gets slower, resulting in fewer vehicles and thus higher speeds on link (14,11). Both cases lead to a negative correlation between (12,14) and (14,11).
      Moreover, the speeds on two neighbouring links in opposite directions could be significantly negatively correlated as well. An example is given by the correlation (-0.39) of travel speeds on link (14, 11) and link (11, 14) in time period 16:00-16:05.
      Similar correlation findings have also been reported in the literature \citep{Cheng2012,Rachtan2013,Ermagun2017,Guo2019}.

  \item Figure~\ref{fig:fig3} exhibits the temporal correlations of link (14, 11) between the first time period and all following time periods in six different 2-hour time ranges, where the stars in dashed lines represent the statistically insignificant correlations with a significance level of 0.05. It is clear that, with an increased time distance between two time periods, the temporal correlation on one link between these two time periods decreases on the whole. And the travel speeds on link (14,11) in close by time periods could have strong positive correlations.
      We also calculated the temporal correlations on other 30 randomly chosen links in the same six 2-hour time ranges. We found only one single (insignificant) negative temporal correlation. These results are in line with the findings of \citet{Cheng2012} and \citet{Rachtan2013}.
\end{enumerate}

We also studied the marginal distributions. It turned out that fitting parametric distributions to the data was not possible due to ``outliers". But these outliers (typically speeds close to zero on roads that normally have more or less free float) are not errors as they do represent real possible speeds. Hence, when we later in this paper need marginal distributions, we use the empirical marginal distributions directly without trying to fit parametric distributions to the data. However, it might be useful to know that when these ``outliers" were removed, the marginal distributions split nicely approximately 50/50 between log-normal and bounded beta distributions.

\section{Scenarios Generation and Solution Evaluation}
\label{sec:SG}
\subsection{Scenarios Generation}

Having the data, we need to generate scenarios for our calculations. We have chosen to approach the problem as follows. We take our data set -- see Section~\ref{sec:case} for details -- and view this empirical distribution as the true distribution of speeds. Obviously, the set is actually a sample from an underlying unknown distribution (or process), but it is beyond the goals of this paper to discuss the relationship between this empirical distribution and the true one. In this way we are in line with the literature \citep{Yang2014,Zockaie2016,Gu2019}.

The base case will be sampling from the empirical distribution. So whatever number of scenarios we need, we shall randomly pick that number of scenarios from the empirical distribution. We shall refer to this as RS (random sampling).

The alternative, which will turn out to be much better, will be based on explicit scenario generation. We shall denote this SG (scenario generation).
We use the method in \cite{Kaut2014} to generate scenarios. This method is based on Sklar's theorem \citep{Sklar73}, showing that every multivariate distribution function is fully specified by its marginal distributions and its copula, a multivariate cdf describing the dependence between the margins. In other words, modelling of the dependence can be fully decoupled from the marginal distributions. However, since multivariate copulas are difficult to work with, the method approximates them by a set of bi-variate copulas.
The scenarios are generated in two steps: first, the method tries to construct a ``scenario copula" as close to the empirical copula as possible, then the scenarios are transformed to the target marginal distributions. Both the copula and marginal distributions can be specified using parametric families, or taken directly from provided data. In our case, the bi-variate copulas and marginal distributions are taken directly from the empirical distribution.

Throughout the text, we refer to correlations. This is to find our place in the literature. But as just mentioned, we are in fact using bi-variate copulas. These are closely related since the only parameters needed for normal copulas are the correlations. However, we are not estimating correlations directly, we are using the empirical copula as a starting point.

The method uses randomness as a tie-breaker if a tie appears in the assignment of this method. This method thus generates the same or very similar scenarios in two different runs with the same input. This contrasts it with sampling where this is not the case. In addition, the means of the scenario set generated by this method are always equal to the means of the given distribution, since we have full control of the discretization. As a result, the mean of travel times on each link and each path are correct in these scenarios, even though they would not reflect all other aspects of the distribution completely.

\subsection{Evaluation Measures and Stability}
The chosen scenarios can have great impacts on the best solution to the SSP problem with spatially and temporally correlated stochastic travel times \citep{Huang2018}. It is thus critical to assess the quality of a set of scenarios. We shall test the qualities when scenarios come from both RS and SG.

We use two related measures, denoted $RD$ and $V\!AR$, to assess the stability / quality of the set of scenarios.
These two measures come from the in- and out-of-sample stability tests of \cite{Michal2007} and work by generating and comparing multiple scenario sets.
However, since the copula-based scenario generation method generates the same or similar scenario sets with the same inputs in different runs, the common in-sample stability test is not applicable here. Thus, we use a variant of the standard approach; to represent the case with $S$ scenarios, we generate $2m+1$ scenario sets with size $S-m, S-m+1,\dots,S,\dots, S+m-1, S+m$, where $m$ is an integer and set as 4 in our paper, based on tests in Guo, Wallace and Kaut \citeyearpar{Guo2018}.
For each scenario set, we take the problem at hand (in our case the SSP problem) and use a solution method from the literature to find an optimal (or near-optimal) solution. Then, there are $2m+1$ best solutions $X_{S+i}$ for $i \in [-m, m]$.
For each such solution, calculate the objective value of $X_{S+i}$ based on each of these $2m+1$ scenario sets.

Let $F^{+}(X_{S+i})$, $F^{-}(X_{S+i})$ and $\sigma_{S+i}$ be the maximum, the minimum and the variance of the $2m+1$ objective values corresponding to $X_{S+i}$.
Then we define the relative difference ($RD$) and variance ($V\!AR$) of the set as follows:

\begin{equation}
RD = \max_{i \in [-m,m]} \left(\frac{F^{+}(X_{S+i})- F^{-}(X_{S+i})}{F^{+}(X_{S+i})}\times100\%\right) \label{eq:Q1} \\
\end{equation}

\begin{equation}
V\!AR = \max_{i \in [-m,m]} \left(\sigma_{S+i}\right) \label{eq:Q2}
\end{equation}

For a given stability requirement, e.g., $RD\leq1\%$ or $RD\leq2\%$, the minimal $S$ satisfying the requirement is set as the number of scenarios necessary to achieve the corresponding objective function stability for the investigated problem.

It is important to understand that by seeing the empirical distribution as the true distribution, we are favouring RS. With a reasonably large number of scenarios, relative to the number of outcomes in the empirical distribution, sampling scenarios from the empirical distribution, without replacement, would look much better than had we sampled from the underlying distribution (or alternatively from the empirical distribution with replacement). The scenario set from sampling without replacement would be ``perfect'' much faster than it would have we sampled from the underlying distribution (or sampled with replacement), while SG would not principally change. We shall see in Sections~\ref{sec:diffODs}--\ref{sec:diffOBJs}, that even so, SG is much stronger than RS.

Let $X_{all}$ be the optimal solution to the investigated SSP problem using all scenarios. We finally use a measure, $ORD$, to check for the performance difference between the optimal solution based on scenarios generated by SG or RS and the optimal solution $X_{all}$ based on all available scenarios. This measure makes sense in our case since we are able to optimally solve our problem with the full empirical distribution, though at considerable computational costs. So even though this test cannot be performed generally, we use it here since it helps us understand what is going on. We might have that, for example, $RD$ and $V\!AR$ both behave well, but the objective function value converges to the wrong value.
Let $F^{all}(X_{all})$ be the optimal objective value when all scenarios are used, and $F^{all}(X_{S+i})$ be the objective value of solution $X_{S+i}$ evaluated with all scenarios (that is, the true value).
$ORD$ is calculated by using Equation~(\ref{eq:Q3}).

\begin{equation}
\label{eq:Q3}
ORD = \frac{1}{2m+1}  \sum_{i=-m}^{m}{\frac{(F^{all}(X_{S+i})-F^{all}(X_{all}))}{F^{all}(X_{all})}}\times 100\% \, \geq 0
\end{equation}

For SG, this is a genuine issue because the method is not guaranteed to be unbiased. We shall see, though, that it behaves very well. The lower $ORD$ is, the less $S$ scenarios produce biased results, provided stability is established.
For RS, the performance difference between $F^{all}(X_{S+i})$ and $F^{all}(X_{all})$ is only caused by sampling errors since random sampling is unbiased.
To reduce the impacts of uncertainty coming from RS itself, we report the minimum, average and maximal values of $RD$ and $V\!AR$ in ten runs, each using $2m+1$ scenario sets (the same number as for SG), as the results for RS in Sections~\ref{sec:diffODs}--\ref{sec:diffOBJs}.

\section{Test Case and Setting}
\label{sec:case}
Extensive experiments are conducted to investigate the stability of objective function evaluations using RS and SG and compare the performances of the two methods for investigated SSP problems. In these experiments, the SSP problem instances are defined on the freeway network around Los Angeles as shown in Figure~\ref{fig:fig5}, and the same 102-day real travel speed data described in Section~\ref{sec:stochasticity} are used.
There are 10,512 random speed variables and 55,245,816 distinct correlations in each 2-hour time range. When using the copula-based scenario generation method, we obtain the marginal distribution of each random speed variable and the 55,245,816 distinct bivariate copulas of all random variables based on the 102-day speed data.

We use the method proposed by \cite{Hall1986} to calculate the shortest paths in stochastic spatially and temporally correlated networks. This method is originally proposed to find the minimum expected travel time path on a stochastic time-dependent network, and we modify it to find the optimal paths for SSP problems with different objective functions.
Specifically, this method is an improvement heuristic, which integrates the branch-and-bound method and the K-shortest path technique to iteratively look for feasible paths until the optimal path is obtained.
At each iteration $k$, the method explores a new path $P_k$ that has the $k^{th}$ minimum possible travel time $g(P_k)$ using a K-shortest path method -- Yen's method \citep{Yen1971} in this paper, and then calculate the actual expected travel time $f(P_{k-1})$ of path $P_{k-1}$. $f(P_{k-1})$ is the expected value of $P_{k-1}$'s travel times in all $S$ scenarios.
Let $\tau$ denote the minimum expected travel time of all paths evaluated already.
We have $\tau = \min\{f(P_1), \dots, f(P_{k-1})\}$.
Through exploring new paths iteratively, the optimal path with the minimum expected travel time can be found when meeting $\tau < g(P_k)$.
The proof is straightforward. For $i \geq 1$, we have $g(P_k)\leq g(P_{k+i})$ and $g(P_i) \leq f(P_i)$ according to the K-shortest path method as well as the definitions of $g(P_i)$ and $f(P_{i})$.
If $\tau < g(P_k)$, $\tau$ is less than the minimum possible travel times $g(P_{k+i})$ of all paths not evaluated yet, which is thus less than the actual expected travel time of all these paths.
Consequently, $\tau$ is the globally minimum expected travel time and its corresponding path is the optimal path.
The method is a general framework to solve optimally the fast path problem with time-dependent stochastic travel times. This framework can be easily modified to solve SSP problems with other objectives by adapting the method of calculating $g(P_k)$  to different objective functions.
Details of the original and modified methods can be found in Appendix A. Of course, other shortest path methods can also be used, a survey of which can be found in \cite{Gendreau2015} and \cite{Madkour2017}. However, the shortest path method is not the focus of this paper.

Sections~\ref{sec:diffODs} and \ref{sec:diffOBJs} will present the stability results for RS and SG from different perspectives, such as the effects of different O-D pairs and different objective functions, respectively, on the stability of function evaluations. The experiments were performed on a laptop with an Intel Core i7-8550U CPU @2.00GHz and 16 GB RAM. All algorithms were coded and executed in MATLAB 2016b.

\section{Effects of Different Origin-Destination Pairs}
\label{sec:diffODs}
We examine and compare the effects of different O-D pairs on the stability and performances of RS and SG first, based on 32 different O-D pairs. Here we only present the results of twelve representative O-D pairs on the basis of different O-D distances and whether the path passes busy traffic areas. Details of the chosen O-D pairs are presented in Table~\ref{tab:12ODs}. The other twenty O-D pairs lead to similar results.

\begin{table}
\caption{Details of the 12 chosen O-D pairs.}
{\begin{tabular}{ccl}\toprule  
No.& O-D pair                 & Locations \\ \midrule
1           & (65, 35)                 & Busy traffic area to common area with long distance  \\
2           & (19, 150)                & Busy traffic area to common area with long distance \\
3           & (16, 29)                 & Busy traffic area to common area with short distance  \\
4           & (16, 26)                 & Busy traffic area to common area with shorter distance \\
5           & (38, 15)                 & Common area to busy traffic area with long distance  \\
6           & (37, 66)                 & Common area to busy traffic area with long distance  \\
7           & (41, 82)                 & Common area to common area with long distance passing busy traffic area \\
8           & (152, 85)                & Common area to common area with long distance not passing busy traffic area \\
9           & (22, 42)                 & Common area to common area with long distance not passing busy traffic area \\
10          & (57, 6)                  & Common area to common area with short distance passing busy traffic area \\
11          & (84, 119)                & Common area to common area with short distance passing busy traffic area \\
12          & (36, 72)                 & Common area to common area with short distance not passing busy traffic area \\ \bottomrule
\end{tabular}}
\label{tab:12ODs}
\end{table}

In these experiments, the departure time is set to 8am during the morning rush hours. The objective function $F1$ is to minimize a linear combination of mean ($\mu$) and standard deviation ($\sigma$) of path travel times, and this problem is referred as the mean--standard deviation shortest path problem \citep{Zhang2017}. Let $\theta$ (generally larger than zero) denote a specified weight factor that represents the risk aversion to travel time variability, $T^{s}$ the travel time (unit: second) of the travel path in the $s^{th}$ scenario, and $S$ the total number of scenarios. This objective (Objective Function 1) is formulated as follows.

\begin{equation}
\label{eq:F1}
 F1 = \mu + \theta \!\cdot\! \sigma
\end{equation}
with $\mu = \frac{\sum_{s=1}^{S}{T^{s}}}{S}$ and $ \sigma = \sqrt{\frac{\sum_{s=1}^{S}(T^{s}-\mu)^2}{S}}$.

\begin{sidewaysfigure}
\centering
\resizebox*{22cm}{!}{\includegraphics{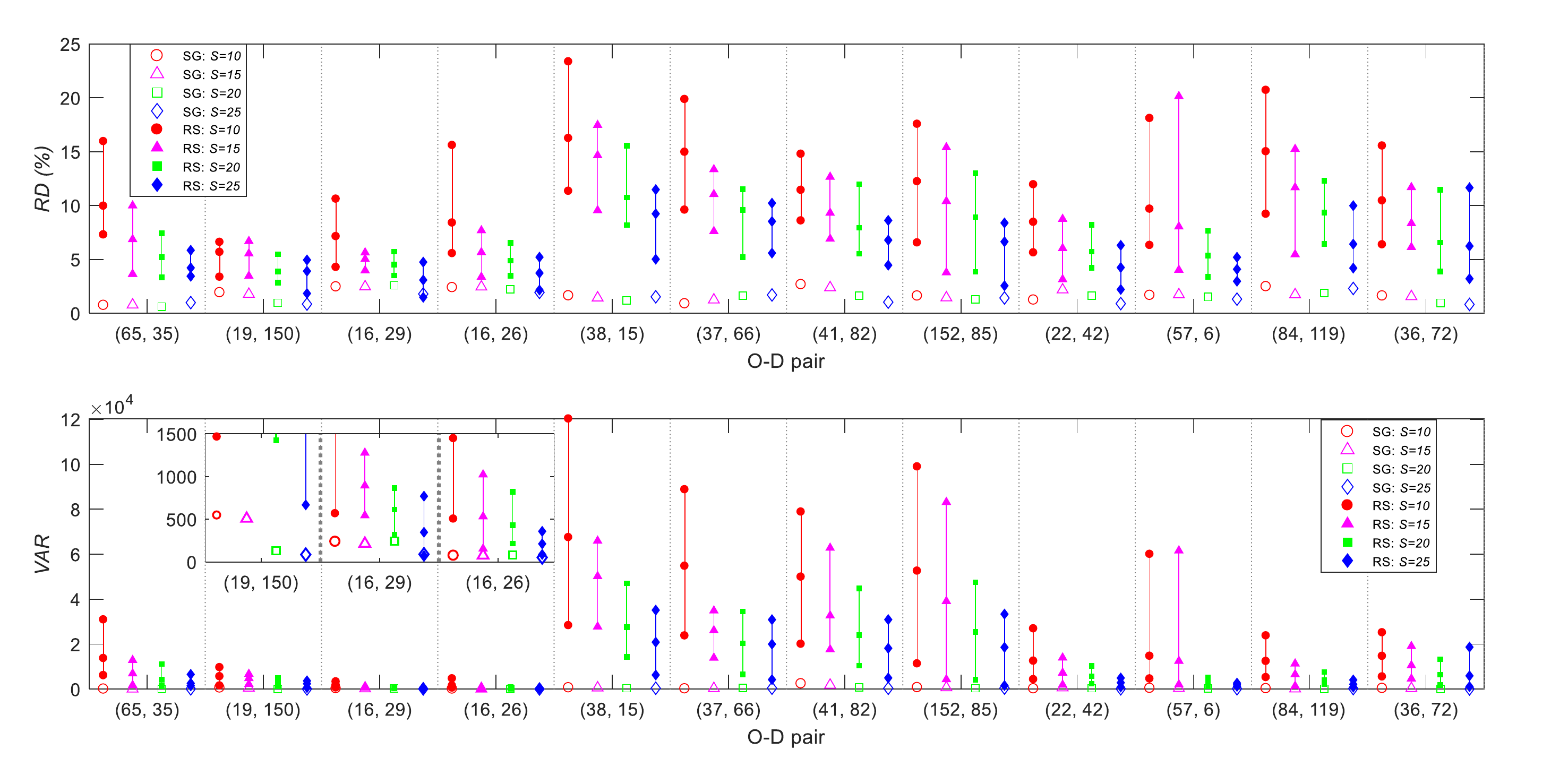}}
\caption{Comparison of $RD$ and $V\!AR$ as generated by RS and SG for different O-D pairs. The three markers on each vertical line, from top to bottom, represent the maximum, the mean, and the minimum of $RD$s or $V\!AR$s generated by RS in ten runs at a certain $S$.}
\label{fig:diffODs}
\end{sidewaysfigure}

\begin{figure}
\centering
\resizebox*{13cm}{!}{\includegraphics{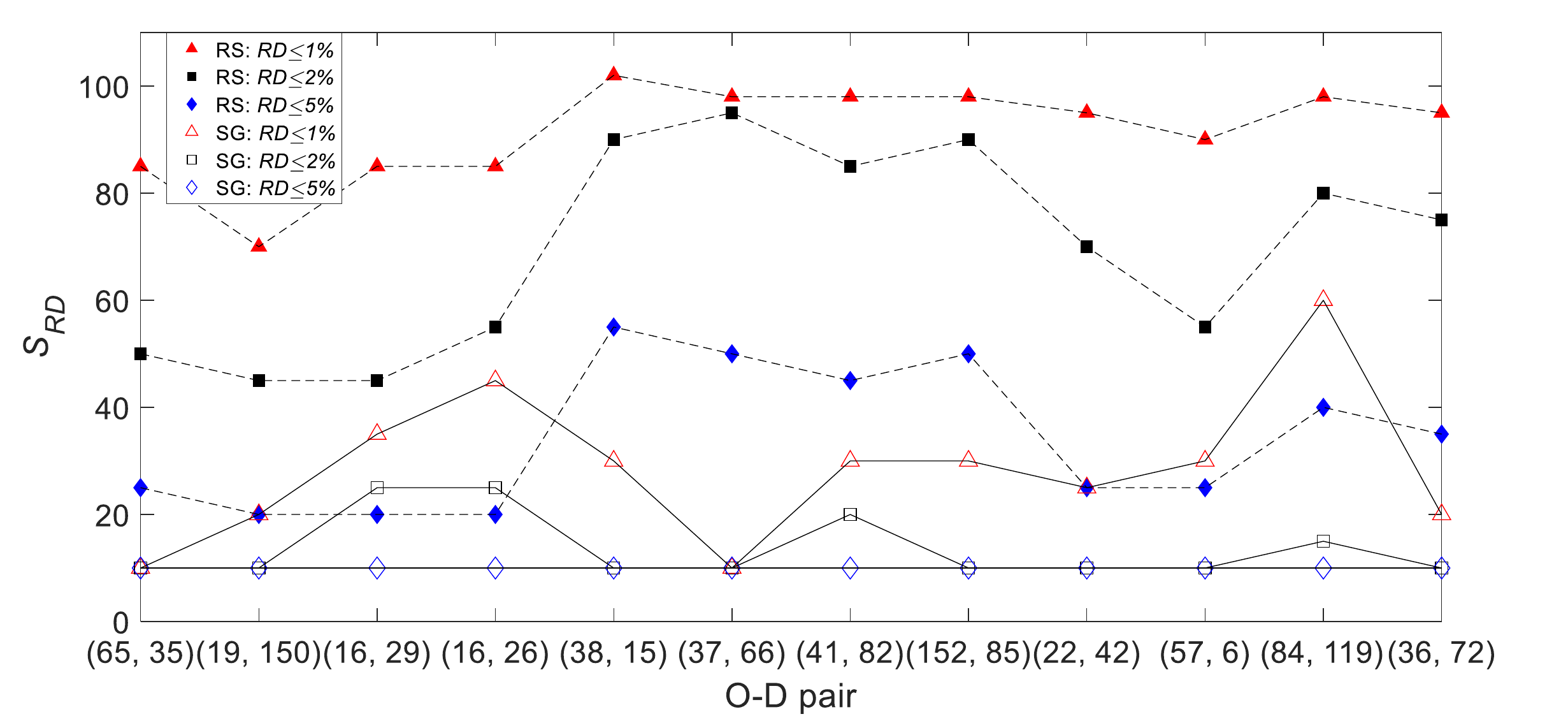}}
\caption{Number of scenarios required by RS and SG for different O-D pairs and different stability levels.}
\label{fig:S_diffODs}
\end{figure}

Figure~\ref{fig:diffODs} compares $RD$s and $V\!ARs$ for the twelve O-D pairs when $\theta$ is set to a typical value 1.27 \citep{Noland1998}. For each O-D pair, the four types of hollow (solid) markers represent the results of SG (RS) at four different values of $S$. Specifically, the three markers on each vertical line, from top to bottom, represent the maximum, the mean, and the minimum of $RD$s or $V\!AR$s generated by RS in ten runs at a certain $S$. The numerical values of $RD$s and $VAR$s in Figure~\ref{fig:diffODs} are presented in Table B.1 of Appendix B.
Table~\ref{tab:ORD_diffODs} presents the $ORD$ results.
Figure~\ref{fig:S_diffODs} shows the number of scenarios required ($S_{RD}$) for different O-D pairs to achieve the specified $RD$ goals for both methods.
For RS, the number of scenarios needed is calculated based on the mean of $RD$ values in ten runs (this also favours RS as compared to SG, since in a given setting one could end up with any of the ten cases without knowing the actual quality of the scenarios used). We let $S$ start from 10 and increase in steps of 5.

\begin{table}
\centering
\caption{$ORD$(\%) values at different $S$ generated by RS and SG for different O-D pairs.}
\resizebox*{8cm}{!}{{\begin{tabular}{cccccc} \toprule																			
\multirow{2}{*}{No.}     & \multirow{2}{*}{O-D pair}  & $S=10$ & $S=15$ & $S=20$ & $S=25$  \\											
& & RS/SG  & RS/SG  & RS/SG  & RS/SG  \\ \midrule						
1&	(65, 35)&	0.86/0&	0.29/0&	0.31/0&	0.18/0\\
2&	(19, 150)&	0.62/0.50&	0.43/0.13&	0.35/0&	0.35/0\\
3&	(16, 29)&	0.23/0.26&	0.08/0&	0.10/0&	0/0\\
4&	(16, 26)&	0.11/0&	0/0&	0/0&	0/0\\
5&	(38, 15)&	0/0&	0/0&	0/0&	0/0\\
6&	(37, 66)&	0.61/0&	0.30/0&	0.28/0&	0.36/0\\
7&	(41, 82)&	0.38/0.08&	0.22/0.07&	0.20/0.09&	0.16/0.12\\
8&	(152, 85)&	1.34/0&	0.58/0&	0.15/0&	0.13/0\\
9&	(22, 42)&	0.07/0&	0/0&	0/0&	0/0\\
10&	(57, 6)&	0.23/0&	0.26/0&	0/0&	0/0\\
11&	(84, 119)&	0.56/0&	0.36/0&	0.38/0&	0.33/0\\
12&	(36, 72)&	0.32/0&	0.08/0&	0/0&	0/0\\
\bottomrule												
\end{tabular}}}
\label{tab:ORD_diffODs}								
\end{table}

It can be found from Figures~\ref{fig:diffODs}-\ref{fig:S_diffODs} and Table~\ref{tab:ORD_diffODs} that:
\begin{enumerate}
  \item Whichever method is used, the $RD$s and $V\!ARs$ reduce with the increase of $S$ on the whole, although small fluctuations exist for some O-D pairs. These fluctuations are reasonable and acceptable, and are caused by the randomness or heuristic nature of RS and SG.

  \item For all O-D pairs, SG generates much smaller $RD$s and $V\!ARs$ than RS, even for its minimal $RD$s and $V\!ARs$. Taking O-D pair (37, 66) as an example, when $S$ is equal to 10, the $RD$ and $V\!AR$ generated by SG are only about 9.23\% and 0.72\% of the minimal $RD$ and $V\!AR$ generated by RS. That is, with the same $S$ (when it is not too large), the scenario set generated by SG leads to much more stable performance than RS for all investigated O-D pairs.

  \item We see a few cases of $ORD$ being positive for SG. However, except for (41,82), it becomes zero as $S$ increases within Table~\ref{tab:ORD_diffODs}. For the O-D pair (41,82), $ORD$ goes to zero when $S$ reaches 55 (not shown in the table). Hence, we see that SG does not produce biased results in our experiments, and the variation is caused by a not-yet-stable solution. We see throughout that $ORD$ is much larger for RS than for SG.

  \item As shown in Figure~\ref{fig:S_diffODs}, to achieve a stability level of 2\% on objective $F1$, RS needs only 25 scenarios for all O-D pairs while the number of scenarios required by RS ranges from 45 to 95 (those are very large numbers as their maximum is 102).

\end{enumerate}

\section{Effects of Different Objective Functions}
\label{sec:diffOBJs}
\subsection{Results of Six Objective Functions}
\label{sec:ResultsOf6Objs}
We further examine the effects of different objective functions on the performances of RS and SG by comparing these two methods based on six commonly used objective functions for all investigated O-D pairs. The departure time is set at 8am. The first objective is $F1$ formulated in Section~\ref{sec:diffODs} and the other five are described below. Objective functions $F1$ and $F6$ are travel time reliability criteria adopted in the literature.

Objective 2: Minimize the expected travel time $F2$.
\begin{equation}
\label{eq:F2}
  F2 = \frac{\sum_{s=1}^{S}{T^{s}}}{S}
\end{equation}

Objective 3: Minimize the expected carbon emission $F3$. Let $e^{s}$ denote the carbon emission (unit: gram) of the travel path in the $s^{th}$ scenario. The value $e^{s}$ is calculated using the well-known MEET model \citep{Hickman1999} with parameters for a gross vehicle weight of 3.5-7.5 tons, i.e., parameters $K$, $a$, $b$, $c$, $d$, $e$, and $f$ are equal to 110, 0, 0, 0.000375, 8702, 0 and 0, respectively. $N$ is the number of nodes in the network, and $M$ is the number of time periods, with $v^{s}_{ijt}$ representing the travel speed sample on link ($i$, $j$) in the $t^{th}$ time period in the $s^{th}$ scenario, and $d_{ij}$ is the length of link ($i$, $j$). If the vehicle travels from node $i$ to node $j$ directly, $x_{ij}$ is equal to 1, otherwise it is 0.
\begin{equation}
\label{eq:F3}
 F3 = \frac{\sum_{s=1}^{S}{(e^{s}/1000)}}{S}
\end{equation}
with
\begin{equation}
e^{s}\!=\!\sum_{i=1}^{N}\sum_{j=1}^{N}(\sum_{t=1}^{M}(K + a\!\cdot\! v^{s}_{ijt} + b\!\cdot\!{v^{s}_{ijt}}^{2} + c\!\cdot\!{v^{s}_{ijt}}^{3} + d/v^{s}_{ijt} + e/{v^{s}_{ijt}}^{2} + f/{v^{s}_{ijt}}^{3}))\!\cdot\! d_{ijt} \!\cdot\! x_{ijt},\nonumber\\
\forall i\neq j, s=1,\dots,S\nonumber
\end{equation}

Objective 4: Minimize the expected tardiness $F4$. Let $L^{s}$ represent the tardiness in the $s^{th}$ scenario, with $D$ the given due time of arriving at the destination node and $D_p$ the departure time from the origin node.
\begin{equation}
\label{eq:F4}
 F4 = \frac{\sum_{s=1}^{S}{L^{s}}}{S}
\end{equation}
with $L^{s}=\max(D_p+T^{s}-D, 0)$.

Objective 5: Minimize the expected sum of tardiness and earliness $F5$. Here $W^{s}$ represents the earliness in the $s^{th}$ scenario, while $E$ is the given earliest time of arriving at the destination node.
\begin{equation}
\label{eq:F5}
 F5 = \frac{\sum_{s=1}^{S}{(L^{s}+W^{s})}}{S}
\end{equation}
with $L^{s}=\max(D_p+T^{s}-D, 0)$ , and $W^{s}=\max(E-T^{s}-D_p, 0)$.

Objective 6: Minimize the travel time budget $F6$ for a specified on-time arrival probability $\alpha$, which is referred as the $\alpha$-reliable path problem \citep{Chen2018} or the minimal percentile travel time path problem \citep{Yang2017}. A risk-averse decision-maker would prefer a larger $\alpha$ for a more reliable path.

\begin{equation}
\label{eq:F6}
 F6 = \overline{T}
\end{equation}
s.t. \begin{equation}
Pr\{T^{s} \leq \overline{T}\} \geq \alpha \nonumber
\end{equation}

Specifically, based on discrete scenarios, the calculation of $F6$ for a path is conducted as follows. First, the travel time in each scenario ($T^s$) is calculated and ranked in an increasing sequence as $T^{s'_1} \leq T^{s'_2} \leq \dots \leq T^{s'_S}$. Then, the value of $F6$ is equal to $T^{\overline{s'}}$, where $\overline{s'} = \min\{s'|\sum_{i=1}^{s'}p_{i}\geq \alpha\}$ and $p_i$ is the possibility of $i^{th}$ scenario. That is, the value of $F6$ is equal to the travel time in the $\overline{s'}^{th}$ scenario \citep{Yang2017}.

\begin{sidewaysfigure}
\centering
\includegraphics[width=\textwidth]{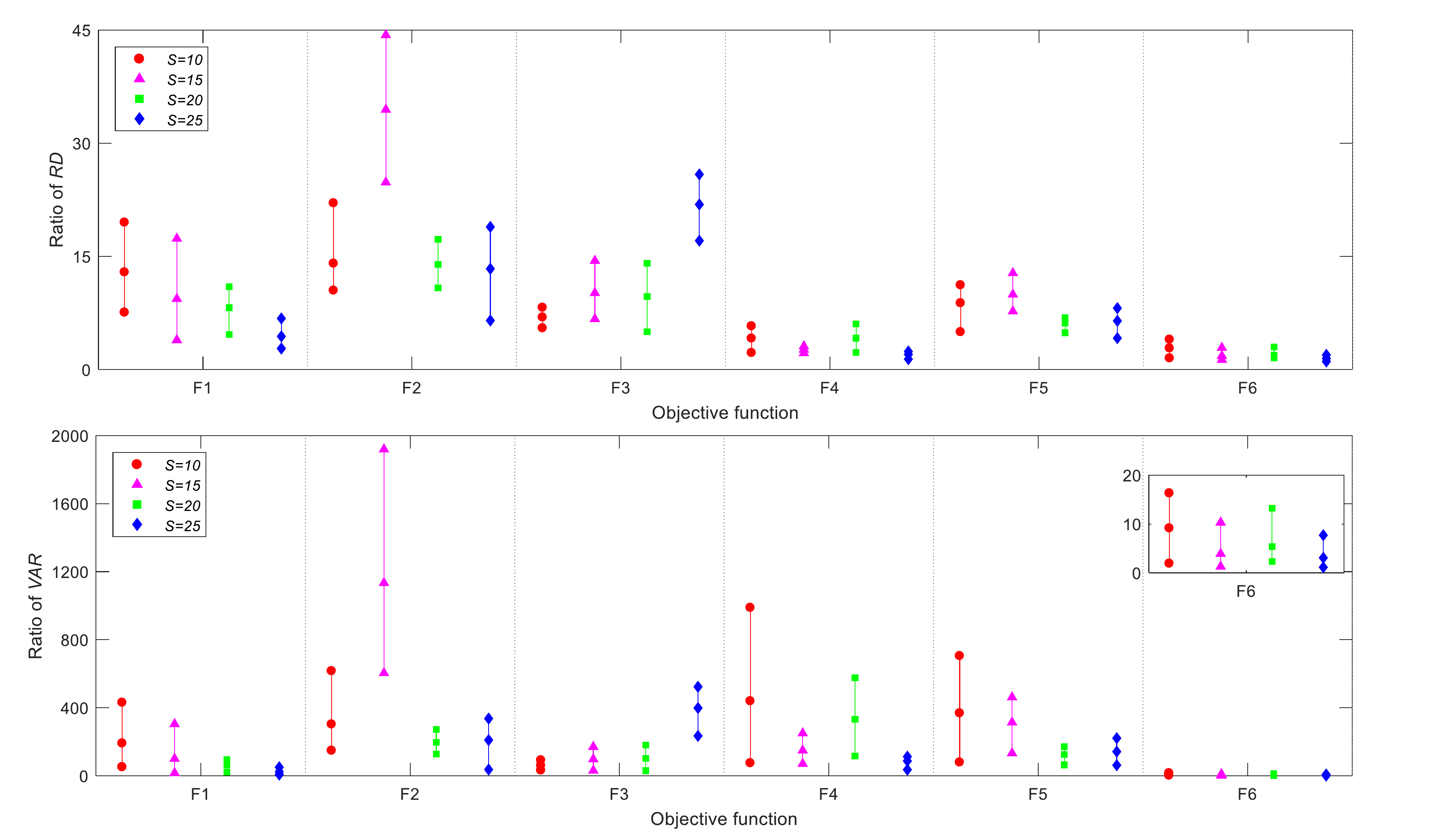}
\caption{Ratios of $RD$ and $V\!AR$ generated by RS and SG at different $S$ for O-D pair (65, 35) and different objective functions. The three markers on each vertical line, from top to bottom, represent the maximum, the mean, and the minimum of $RD$ or $V\!AR$ ratios in ten runs at a certain $S$.}
\label{fig:diffobjs}
\end{sidewaysfigure}

To compare the results of different objective functions, we take O-D pair (65, 35) and departure time $D_p$ at 8am as an example. Figure~\ref{fig:diffobjs} shows the ratios of $RD$ and $V\!AR$ generated by RS and SG for different $S$. The ratio of $RD$ ($V\!AR$) is defined as the $RD$s ($V\!AR$s) generated by RS divided by the corresponding $RD$s or $VAR$s generated by SG.
The three markers on each vertical line, from top to bottom, represent the maximum, the mean, and the minimum of $RD$ or $V\!AR$ ratios in ten runs at a certain $S$.
The numerical values of $RD$s and $VAR$s in Figure~\ref{fig:diffobjs} are presented in Table B.2 of Appendix B.
Figure~\ref{fig:S_diffobjs} presents the number of scenarios required in each case.
We set $\theta$ to 1 in $F1$, due time $D$ to 8.88 (i.e., 8:52:48 in HH:MM:SS format) in $F4$, time window ($E$, $D$) to (8.78, 8.87) in $F5$, and on-time arrival probability $\alpha$ to 90\% in $F6$. We set $\theta=1$ in $F1$ so that the results of $F1$ here are consistent with the results of different $\theta$ shown in Table~\ref{tab:F1}, which corresponds to the experiments of studying the effects of different $\theta$ values in Section~\ref{sec:eff_parameters}. Other O-D pairs produce similar results.

\begin{figure}
	\centering
	\includegraphics[width=\textwidth]{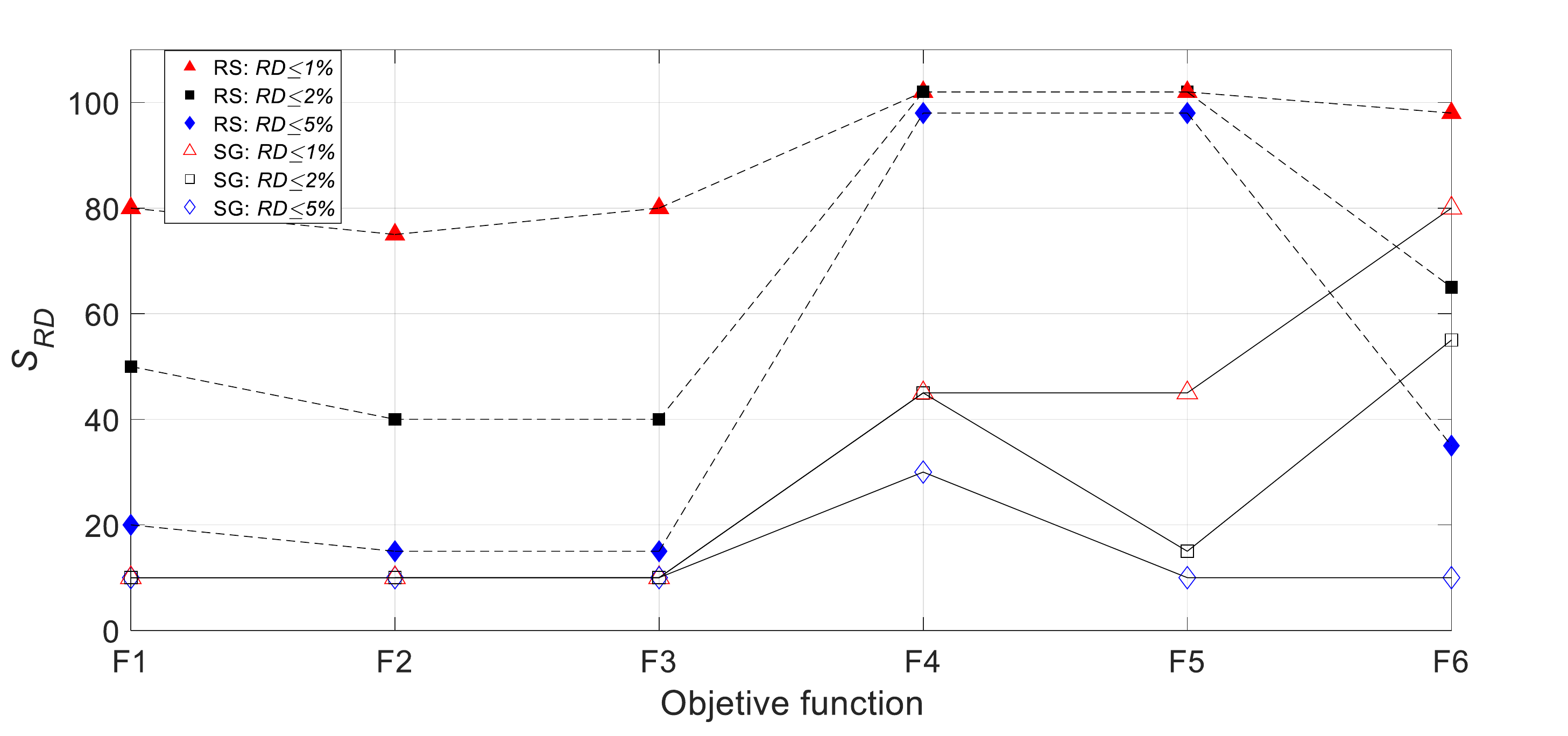}
	\caption{Number of scenarios required by RS and SG for O-D pair (65, 35), when different objective functions and different stability levels are considered.}
\label{fig:S_diffobjs}
\end{figure}

It can be found from Figures~\ref{fig:diffobjs}-\ref{fig:S_diffobjs} that:
\begin{enumerate}
  \item The objective function matters for the ratios for $RD$ and $V\!AR$. We see that the ratios can be substantial for many objective functions, so that SG produces much more stable results than RS. The observed values across objective functions are not directly comparable, but still indicates well the dangers of using pure sampling.

\item To achieve specified $RD$ goals, different objective functions require different numbers of scenarios.
      For objective functions $F2$ and $F3$ that calculate means, to achieve $RD\leq1\%$, SG needs only 10 scenarios while RS requires more than or equal to 75 scenarios, and more similar results can be found in Appendix B.
      For the objective functions $F4$ and $F5$ that consider earliness and / or tardiness, both SG and RS need many scenarios, however, SG needs fewer scenarios to achieve $RD\leq1\%$ than RS needs to achieves $RD\leq5\%$.

\end{enumerate}

\subsection{Effects of Parameters in Different Objective Functions}
\label{sec:eff_parameters}
Since the objective functions $F1$, $F4$, $F5$ and $F6$ are directly affected by certain parameter values, we now study the effects of these parameters on $RD$, $V\!AR$ and $ORD$ values.
We conduct experiments for all O-D pairs. For each O-D pair, we customize the values of parameters to investigate as many settings as possible, and then obtain some general observations from the experimental results.

Take O-D pair (65, 35) as an example. The details of the experimental results are shown in Appendix C. We have the following observations:

\begin{enumerate}
  \item $RD$ and $V\!AR$ generated by RS are much larger than those generated by SG throughout.
      This can be seen from Table~\ref{tab:R_RD}, which presents the ranges of the ratio of $RD$ and $V\!AR$ under different parameters for O-D pair (65, 35) and $S=10$.
      In the table, the minimal ratios of $RD$ and $V\!AR$ of all objective functions are larger than 1.
      This indicates that the scenarios generated by SG have more stable performance than those generated by RS for all these objective functions and parameters.

  \item The values of $RD$ and $V\!AR$ can vary greatly in cases with different parameters. The results are not too surprising. When the objective is some kind of expectation, variation is much smaller than when the objective function depends on extreme values of the random variables. When extreme values are important, both methods, but particularly RS, will be very sensitive to what happens to be the extreme values in a scenario set. Detailed analyses can be found in Appendix C.

  \item In most cases, we have $ORD=0$. When it is not the case in the tables, the reason is that more scenarios are needed for stability. Eventually, all $ORD$s go to zero in our tests. In particular, that means that we never observe bias for SG --- which generally is a possibility.

\end{enumerate}

\begin{table}
\caption{Ranges of the ratio of $RD$ and $V\!AR$ generated by RS and SG with different parameters for O-D pair (65, 35) and $S=10$.}
{\begin{tabular}{ccccc} \toprule																			
 & $F1$ & $F4$ & $F5$ & $F6$  \\	\midrule  										
Ratio of $RD$   & {[}8.70, 20.51{]}     & {[}2.25, 11.93{]}        & {[}1.44, 11.72{]}   & {[}1.29, 4.64{]}    \\
Ratio of $V\!AR$  & {[}74.90, 393.06{]} & {[}105.82, 417.60{]} & {[}24.60, 444.80{]} & {[}1.36, 27.25{]}   \\ \bottomrule					
\end{tabular}}
\label{tab:R_RD}						
\end{table}

\par
In summary, to examine the stability / quality of objective function evaluations of RS and SG and compare the performances of these two methods for the investigated SSP problems, Sections~\ref{sec:diffODs} and \ref{sec:diffOBJs} have presented extensive experimental and comparison results in terms of different O-D pairs and different objective functions. The above experimental results show that \par

\begin{enumerate}
\item SG strongly outperforms RS in terms of stability, i.e., in terms of $RD$ and $V\!AR$;
\item Different O-D pairs, as well as different objective functions and parameters could have large effects on the stability of objective function evaluations;
\item For objective functions $F2$ and $F3$, SG needs only 10 scenarios for almost all cases to achieve an objective-function evaluation stability of 1\% while RS usually needs much more;
\item For the other objective functions, sometimes much more scenarios are required to achieve a stability level of 1\% because different parameter values could have a great impact on the $RD$ results. However, SG always needs fewer scenarios than RS.
\item The scenarios generated by SG are unbiased ($ORD=0$ when $S$ is large enough).
\end{enumerate}

\section{Conclusions}
\label{sec:conclusions}

This paper addresses SSP problems with spatially and temporally correlated travel speeds based on real data from a California freeway network. These problems involve correlated and very high-dimensional random speed variables.

We investigate realistic stochasticity of travel speeds in a freeway network with 10,512 correlated random variables and 55,245,816 distinct correlations (or distinct bi-variate copulas). We find that the speeds on neighbouring links and in close by time periods are strongly spatio-temporally correlated.

We then present two methods for generating scenarios for these high-dimen\-sional SSP problems based on realistic stochasticity of travel speeds, one a copula-based scenario generation (SG) method and one a random sampling (RS) method. Extensive experiments are conducted to compare the performances of the two methods by examining the effects of different origin-destination pairs and different objective functions. In terms of two performance measures ($RD$ and $V\!AR$), we show that SG needs much fewer scenarios than RS to achieve the same stability, typically about 6-10 times less for a stability level of 1\% on objective function evaluations. So, our main conclusion is that it is crucial how scenarios are generated, and in particular that random sampling is not a good idea unless the problem is small or execution time is of little importance. The suggested way of generating scenarios requires offline computations. We typically needed 28 minutes for 10 scenarios, up to 50 minutes for 25 scenarios when we had 10,512 random variables. The scenarios can then be reused as long as the map does not change.

This research is conducted based on a freeway network, in which the travel speeds on many links have relatively low variances. Besides, this paper adopts the Pearson correlations of travel speeds to represent their correlations. Our future research will extend the scenario generation method to urban street networks with higher speed variances and more path choices, and further consider partial correlations and partial autocorrelations. Another interesting direction is to investigate the effects of different correlations on the solutions to SSP problems.

%
%
%

%

%

%


\bibliographystyle{apacite}
\bibliography{references}

\clearpage
\appendix

\section{Stochastic Shortest Path Solving Method}
\label{sec:HallMethod}
\setcounter{table}{0}
\renewcommand{\thetable}{A.\arabic{table}}

We use the method proposed by \cite{Hall1986} to handle the SSP problems with different objective functions in stochastic spatially
and temporally correlated networks. This method is originally proposed to find the minimal expected travel time path on a stochastic time-dependent network, and we modify it to find the optimal paths for our SSP problems with different objective functions.

Table~\ref{tab:Hall} shows the framework of Hall's method.
At each iteration $k$, a new solution $P_k$ (i.e., the path with the $k^{th}$ smallest possible objective value) is found using a K-shortest path method -- Yen's method \citep{Yen1971} in this paper, and the actual objective value ($\tau$) of the best possible path is updated.
$\tau$ is the minimum of objective values of all $k-1$ paths evaluated already. That is, $\tau = \min\{f(P_1), \dots, f(P_{k-1})\}$.
Note that the method probably needs to explore and evaluate a large number of solutions to get the optimal one. However, the efficiency of this method is not the concern of this paper.

Next, we verify the optimality of the framework of Hall's method.
As shown in Table~\ref{tab:Hall}, $g(P_k)$ is the minimum possible objective value of path $P_k$, and $f(P_k)$ is the actual objective value of path $P_k$. We have $g(P_k)\leq f(P_k)$.
When $\tau < g(P_m)$ at iteration $m$, we have $ \tau < g(P_m) \leq f(P_m)$, that is, $\tau < f(P_m)$.
Since $g(P_k) \leq g(P_{k+i})$ for all $i\geq 1$, we have $\tau < g(P_m) \leq g(P_{m+i}) \leq f(P_{m+i})$.
That is, $\tau < f(P_{m+i})$ for all $i \geq 1$. Therefore, $\tau$ is not only the minimum objective value of all evaluated paths, but also less than the objective values of all unevaluated paths $P_{m+i}$ ($i\geq0$). $\tau$ is thus the global optimal objective value.

\begin{table}
\centering
\caption{Framework of Hall's method (1986).}
\resizebox*{12cm}{!}{{\begin{tabular}{ll}\toprule
Step No. & DO \\ \midrule
1 & Set $k = 1$, $\tau = \infty$ \\
  & Find the shortest path $P_1$ with the minimum possible objective value $g(P_1)$ \\
2 & Set $k = k + 1$ \\
  & Calculate the actual objective value $f(P_{k-1})$ of $P_{k-1}$ \\
3 & Find the path $P_k$ with the $k^{th}$ minimum possible objective value $g(P_k)$ \\
4 & If $f(P_{k-1}) < \tau$ \\
  & \quad $\tau = f(P_{k-1})$ \\
  & \quad $P^* = P_{k-1}$ \\
  & If $\tau \geq g(P_k)$ \\
  & \quad Return to Step 2 \\
  & If $\tau < g(P_k)$ \\
  & \quad $\tau$ is the optimal objective value \\
  & \quad $P^*$ is the optimal path \\
  & \quad Stop and return $\tau$, $P^*$ \\\bottomrule
\end{tabular}}}
\label{tab:Hall}
\end{table}

In summary, to get the optimal path in this framework, two conditions need to be satisfied: (1) $g(P_k) \leq g(P_{k+1})$ and (2) $g(P_k)\leq f(P_k)$.
The first condition can be easily satisfied by using the Yen's method \citep{Yen1971} to find the K-shortest paths optimally in a static and deterministic network.
To meet the second condition, the key is to set the minimum possible objective value on each road link, which has not been illustrated clearly in Hall's paper (1986).
We set it for our SSP problems with different objective functions as follows. Since the method for $F2$ is the basis for other objective functions, we introduce it first.

\textbf{$F2$: Minimize the expected travel time}

We first find the maximum speed $v_{ij}^{max}$ on each link over all 24 time periods (2-hour time range) and all $S$ scenarios, and then set the value of $d_{ij}/v_{ij}^{max}$ as the minimum possible travel time on each link.

Because $v_{ij}^{max}$ is the highest possible speed on link ($i, j$), $d_{ij}/v_{ij}^{max}$ must be less than or equal to the travel time on this link in any time periods and any scenarios.
Therefore, the sum $g(P_k)$ of the minimum possible travel time on all links of path $P_k$ must be less than or equal to the expected travel time $f(P_k)$ according to Equation~(\ref{eq:F2}). This is, condition (2) holds.

\textbf{$F1$: Minimize the linear combination of mean ($\mu$) and standard deviation ($\sigma$) of path travel times}

According to Equations~(\ref{eq:F1})-(\ref{eq:F2}), for path $P_k$, its objective value of $F1$ must be greater than or equal to the objective value of $F2$ since $\theta$ is generally equal to or bigger than 0.
Therefore, we can directly use the minimum possible travel time for $F2$ as the minimum possible objective value for $F1$, which guarantees condition (2) holds.

\textbf{$F3$: Minimize the expected carbon emissions}

The minimum carbon emission value $e_{ij}^{min}$ over all time periods and all scenarios is set as the minimum possible carbon emission on the link.
According to Equation~(\ref{eq:F3}), the sum $g(P_k)$ of the values of $e_{ij}^{min}$ on all links of path $P_k$ must be less than or equal to the expected carbon emission $f(P_k)$ because  $e_{ij}^{min}$ is less than or equal to the carbon emission values on link ($i, j$) in any time periods and any scenarios. Condition (2) holds.

\textbf{$F4$: Minimize the expected tardiness}

To get the minimum possible tardiness for path $P_k$, the vehicle must travel each link at the maximum speed, which is consistent with objective function $F2$. Therefore, we use the same method for $F2$ to get the minimum possible travel time $T^{min}(P_k)$ of path $P_k$, and then the minimum possible tardiness $g(P_k)$ is set as $max(D_p+T^{min}(P_k)-D,0)$.

According to Equation~(\ref{eq:F4}), $g(P_k)$ must be less than or equal to the expected tardiness $f(P_k)$ because $T^{min}(P_k)$ is the minimum possible travel time of path $P_k$, as stated for $F2$. Condition (2) holds.

\textbf{$F5$: Minimize the expected sum of tardiness and earliness}

Here, we need to get the minimum possible tardiness and earliness respectively.
The minimum possible tardiness $L^{min}(P_k)$ of path $P_k$ is calculated using the same method for objective function $F4$.
The minimum possible earliness $W^{min}(P_k)$ of path $P_k$ can always be zero theoretically, as the vehicle could avoid earliness by lowering down the travel speed or temporary parking. Therefore, we set the $W^{min}(P_k)$ of path $P_k$ as zero.

The sum $g(P_k)$ of $L^{min}(P_k)$ and $W^{min}(P_k)$ must be less than or equal to the expected tardiness and earliness $f(P_k)$ according to Equation~(\ref{eq:F5}). Condition (2) holds.

\textbf{$F6$: Minimize the travel time budget for a specified on-time arrival probability $\alpha$}

As stated in Section~\ref{sec:ResultsOf6Objs}, the objective value of $F6$ equals to the vehicle's travel time $T_s$ in one of the $S$ scenarios.
Here, we use the minimum possible travel time generated for  objective function $F2$ as the minimum possible objective value for $F6$ as well.

According to Equation~(\ref{eq:F6}), the sum $g(P_k)$ of the minimum possible travel times on all links in path $P_k$ must be less than or equal to the travel time in any scenarios $f(P_k)$ since the minimum possible travel time $d_{ij} / v_{ij}^{max}$ on link ($i, j$) is less than or equal to the travel time on link ($i, j$) over all time periods and all $S$ scenarios. Condition (2) holds.

\section{Supplementary Result}
\label{sec:Supplementary_result}
\setcounter{table}{0}
\renewcommand{\thetable}{B.\arabic{table}}

In this section, Tables~\ref{tab:Detai_Fig5} and \ref{tab:Detai_Fig7} show the corresponding numerical values of $RD$ and $V\!AR$ results shown in Figures~\ref{fig:diffODs} and \ref{fig:diffobjs}, respectively.
In these tables, the results generated by RS and SG for each $S$ are separated by a slash. The three values for RS in each cell represent the minimum, the mean, and the maximum of $RD$s and $V\!AR$s in ten runs at $S$.
Given different O-D pairs, Tables~\ref{tab:S_F2} and \ref{tab:S_F3} present the number of scenarios required to achieve the
specified $RD$ goals for both methods under objective functions $F2$ and $F3$, respectively.

\begin{table}
\caption{Numerical values of $RD$ and $V\!AR$ results shown in Figure~\ref{fig:diffODs}.}
\resizebox{\linewidth}{0.3\linewidth}{	
{\begin{tabular}{ccccccc} \toprule																			
& \multirow{2}{*}{No.} & \multirow{2}{*}{O-D pair}     & $S=10$ & $S=15$ & $S=20$ & $S=25$  \\												
&     &          & RS/SG  & RS/SG  & RS/SG  & RS/SG \\ \midrule											
 \multirow{12}{*}{$RD$(\%)}	&	1&	(65, 35)&	(7.28, 9.95, 15.94)/0.74&	(3.64, 6.87, 10.00)/0.78&	(3.33, 5.21, 7.42)/0.61&	(3.44, 4.21, 5.85)/0.98	\\
&	2&	(19, 150)&	(3.35, 5.66, 6.59)/1.92&	(3.45, 5.54, 6.69)/1.77&	(2.84, 3.88, 5.48)/0.96&	(1.84, 3.91, 4.94)/0.85	\\
&	3&	(16, 29)&	(4.26, 7.12, 10.60)/2.46&	(3.97, 5.03, 5.63)/2.46&	(3.50, 4.50, 5.71)/2.59&	(1.48, 3.08, 4.74)/1.79	\\
&	4&	(16, 26)&	(5.54, 8.38, 15.58)/2.39&	(3.37, 5.65, 7.68)/2.45&	(3.49, 4.87, 6.54)/2.22&	(2.15, 3.73, 5.21)/1.94	\\
&	5&	(38, 15)&	(11.33, 16.23, 23.35)/1.63&	(9.55, 14.66, 17.46)/1.43&	(8.18, 10.75, 15.55)/1.17&	(5.02, 9.23, 11.48)/1.53	\\
&	6&	(37, 66)&	(9.58, 14.95, 19.85)/0.88&	(7.61, 11.05, 13.35)/1.24&	(5.21, 9.59, 11.5)/1.64&	(5.58, 8.52, 10.22)/1.70	\\
&	7&	(41, 82)&	(8.58, 11.42, 14.77)/2.66&	(6.91, 9.31, 12.65)/2.38&	(5.53, 7.94, 11.99)/1.62&	(4.46, 6.79, 8.62)/1.03	\\
&	8&	(152, 85)&	(6.54, 12.22, 17.55)/1.61&	(3.78, 10.40, 15.40)/1.44&	(3.84, 8.93, 12.98)/1.28&	(2.55, 6.64, 8.38)/1.42	\\
&	9&	(22, 42)&	(5.61, 8.45, 11.94)/1.23&	(3.15, 6.03, 8.74)/2.18&	(4.22, 5.72, 8.20)/1.63&	(2.21, 4.25, 6.30)/0.89	\\
&	10&	(57, 6)&	(6.30, 9.67, 18.09)/1.67&	(4.02, 8.04, 20.14)/1.73&	(3.37, 5.36, 7.64)/1.51&	(2.97, 4.08, 5.20)/1.31	\\
&	11&	(84, 119)&	(9.20, 15.00, 20.70)/2.48&	(5.46, 11.67, 15.24)/1.73&	(6.43, 9.35, 12.30)/1.89&	(4.19, 6.42, 9.98)/2.30	\\
&	12&	(36, 72)&	(6.36, 10.44, 15.53)/1.61&	(6.13, 8.35, 11.69)/1.55&	(3.86, 6.55, 11.45)/0.96&	(3.20, 6.23, 11.65)/0.82	\\\midrule
\multirow{12}{*}{$V\!AR$}&	1&	(65, 35)&	(6047, 13658, 30847)/70&	(1724, 6916, 12871)/91&	(1319, 4156, 11125)/49&	(1496, 2589, 6572)/122	\\
&	2&	(19, 150)&	(1464, 5565, 9605)/545&	(2222, 4826, 6721)/507&	(1426, 2684, 4998)/131&	(669, 2425, 3672)/86	\\
&	3&	(16, 29)&	(566, 1791, 3259)/237&	(544, 894, 1278)/214&	(322, 612, 866)/244&	(73, 348, 770)/90	\\
&	4&	(16, 26)&	(503, 1445, 4618)/74&	(155, 532, 1023)/76&	(215, 428, 820)/82&	(88, 212, 358)/53	\\
&	5&	(38, 15)&	(28282, 67444, 120184)/638&	(27798, 50188, 65956)/590&	(14387, 27499, 47051)/418&	(6297, 20887, 35141)/402	\\
&	6&	(37, 66)&	(23704, 54701, 88716)/170&	(13921, 26106, 34894)/340&	(6539, 20393, 34486)/524&	(4230, 20013, 30885)/428	\\
&	7&	(41, 82)&	(20006, 49856, 78798)/2421&	(17655, 32707, 62801)/1823&	(10448, 23981, 44819)/752&	(4974, 18188, 30921)/382	\\
&	8&	(152, 85)&	(11272, 52525, 98880)/717&	(4353, 39106, 83156)/686&	(4209, 25416, 47481)/404&	(1597, 18596, 33372)/468	\\
&	9&	(22, 42)&	(4262, 12487, 26860)/187&	(1933, 7259, 13977)/636&	(2471, 5722, 10292)/415&	(972, 2866, 4955)/154	\\
&	10&	(57, 6)&	(4570, 14668, 59985)/403&	(2057, 12507, 61605)/373&	(1193, 3377, 5210)/287&	(1229, 1888, 2677)/174	\\
&	11&	(84, 119)&	(5180, 12358, 23734)/255&	(1402, 6553, 11332)/144&	(2195, 4051, 7619)/153&	(903, 2184, 4074)/240	\\
&	12&	(36, 72)&	(5466, 14625, 25113)/357&	(4642, 10547, 19070)/322&	(1881, 6426, 13282)/123&	(1123, 5918, 18681)/83	\\\bottomrule	
\end{tabular}}	}
\label{tab:Detai_Fig5}						
\end{table}

\begin{table}
\caption{Numerical values of $RD$ and $V\!AR$ results shown in Figure~\ref{fig:diffobjs}.}
\resizebox{\linewidth}{0.2\linewidth}{	
{\begin{tabular}{cccccc} \toprule																			
&Objective   & $S=10$ & $S=15$ & $S=20$ & $S=25$  \\												
&function    & RS/SG  & RS/SG  & RS/SG  & RS/SG \\ \midrule											
 \multirow{6}{*}{$RD$(\%)}&	$F1$&	(4.52, 7.7, 11.64)/0.60&	(2.51, 6.00, 11.11)/0.64&	(2.44, 4.30, 5.75)/0.52&	(2.26, 3.56, 5.49)/0.81	\\
&	$F2$&	(4.62, 6.20, 9.73)/0.44&	(3.06, 4.25, 5.47)/0.12&	(2.44, 3.14, 3.89)/0.23&	(1.18, 2.42, 3.42)/0.18	\\
&	$F3$&	(4.70, 5.93, 7.03)/0.85&	(2.66, 4.02, 5.71)/0.40&	(1.81, 3.51, 5.10)/0.36&	(2.49, 3.19, 3.77)/0.15	\\
&	$F4$&	(66.33, 82.42, 100)/4.96&	(50.18, 68.31, 89.49)/1.40&	(49.43, 61.14, 77.11)/2.52&	(44.38, 54.41, 64.62)/2.03	\\
&	$F5$&	(42.61, 53.2, 63.03)/4.45&	(24.33, 44.11, 67.52)/1.25&	(31.15, 38.53, 43.61)/2.26&	(24.58, 29.63, 37.58)/1.82	\\
&	$F6$&	(7.09, 13.25, 18.62)/4.67&	(6.33, 8.88, 14.13)/4.88&	(6.02, 7.31, 11.62)/3.87&	(4.66, 6.35, 8.43)/4.33	\\\midrule
\multirow{6}{*}{$V\!AR$}&	$F1$&	(2356, 8720, 19638)/46&	(953, 6065, 18310)/60&	(741, 2258, 3339)/35&	(601, 1908, 4068)/81	\\
&	$F2$&	(2365, 4831, 9820)/16&	(1153, 2163, 3662)/2&	(791, 1206, 1681)/6&	(132, 751, 1204)/4	\\
&	$F3$&	(0.39, 0.73, 1.11)/0.01&	(0.12, 0.38, 0.67)/0&	(0.07, 0.25, 0.45)/0&	(0.13, 0.22, 0.28)/0	\\
&	$F4$&	(784, 2252, 4859)/16&	(316, 1263, 3017)/2&	(282, 759, 1530)/6&	(250, 475, 824)/4	\\
&	$F5$&	(2678, 4448, 7732)/16&	(590, 1806, 3897)/2&	(703, 1369, 2879)/6&	(371, 770, 1371)/4	\\
&	$F6$&	(6558, 31329, 55924)/3430&	(5172, 15713, 41423)/4022&	(3622, 8377, 20615)/1562&	(2438, 6545, 16439)/2135	\\\bottomrule		\end{tabular}}}		
\label{tab:Detai_Fig7}			
\end{table}

\begin{table}
\centering
\caption{Number of scenarios required for different O-D pairs under $F2$.}
{\begin{tabular}{cccccc} \toprule
 \multirow{2}*{No.}  & \multirow{2}*{O-D pair} &	$RD\leq1\%$	&	$RD\leq2\%$	&	$RD\leq5\%$		\\
    &               & RS/SG     & RS/SG   & RS/SG    \\ \midrule
1   &	(65, 35)	&	70/10	&	45/10	&	15/10		\\
2   &	(19, 150)	&	45/10	&	25/10	&	10/10		\\
3	&	(16, 29)	&	90/10	&	45/10	&	10/10		\\
4	&	(16, 26)	&	85/10	&	50/10	&	20/10		\\
5	&	(38, 15)	&	98/10	&	75/10	&	45/10		\\
6	&	(37, 66)	&	90/10	&	60/10	&	25/10		\\
7	&	(41, 82)	&	90/15	&	65/10	&	25/10		\\
8	&	(152, 85)   &	95/10	&	65/10	&	20/10		\\
9	&	(22, 42)	&	90/10	&	45/10	&	10/10		\\
10	&	(57, 6)	    &	90/15	&	55/10	&	25/10		\\
11	&	(84, 119)	&	98/15	&	85/10	&	45/10		\\
12  &   (36, 72)	&	98/10	&	55/10	&	15/10		\\\bottomrule
\end{tabular}}
\label{tab:S_F2}
\end{table}

\begin{table}
\centering
\caption{Number of scenarios required for different O-D pairs under $F3$.}
{\begin{tabular}{cccccc} \toprule
 \multirow{2}*{No.}  & \multirow{2}*{O-D pair}  & $RD\leq1\%$	&	$RD\leq2\%$	&	$RD\leq5\%$		\\
    &               & RS/SG     & RS/SG   & RS/SG    \\ \midrule
1&	(65, 35)&	80/10&	45/10&	15/10	\\
2&	(19, 150)&	65/10&	25/10&	10/10	\\
3&	(16, 29)&	70/10&	40/10&	10/10	\\
4&	(16, 26)&	75/10&	35/10&	10/10	\\
5&	(38, 15)&	98/10&	98/10&	25/10	\\
6&	(37, 66)&	98/10&	98/10&	15/10	\\
7&	(41, 82)&	80/10&	50/10&	15/10	\\
8&	(152, 85)&	75/10&	50/10&	15/10	\\
9&	(22, 42)&	98/10&	98/10&	10/10	\\
10&	(57, 6)&	98/10&	98/10&	20/10	\\
11&	(84, 119)&	85/15&	55/10&	20/10	\\
12&	(36, 72)&	70/10&	30/10&	10/10	\\\bottomrule
\end{tabular}}
\label{tab:S_F3}
\end{table}

\newpage
\section{Detailed Results of O-D Pair (65, 35)}
\label{sec:figure_parameters}
\setcounter{table}{0}
\renewcommand{\thetable}{C.\arabic{table}}

In Section \ref{sec:eff_parameters}, we use O-D pair (65, 35) as an example to describe the effects of different parameters in the objective functions on $RD$, $V\!AR$, and $ORD$ results.
The detailed results of O-D pair (65, 35) under different objective functions are shown in this section.
Tables~\ref{tab:F1}, \ref{tab:F4}, \ref{tab:F5} and \ref{tab:F6} show the $RD$ and $V\!AR$ results at different $S$ for objective functions $F1$, $F4$, $F5$ and $F6$, respectively. The format of these tables
is the same with the format of Tables~\ref{tab:Detai_Fig5} and \ref{tab:Detai_Fig7}.
Tables~\ref{tab:ORD_F1}, \ref{tab:ORD_F4}, \ref{tab:ORD_F5}, and \ref{tab:ORD_F6} show the $ORD$ results for objective functions $F1$, $F4$, $F5$ and $F6$, respectively.
In these tables, the values separated by a slash in each cell represent the results generated by RS and SG, respectively.

To further clarify Point 2 in Section~\ref{sec:eff_parameters}, we give some detailed analyses for it as follows.
\begin{enumerate}
    \item When the objective is some kind of expectation of travel time (i.e., approximates the expected travel time $F2$), the variation represented by $RD$ and $V\!AR$ is relatively small.
    As shown in the tables of $RD$ and $V\!AR$ results, $F1$ with small $\theta$, $F4$ with small $D$, $F5$ with small time interval $D-E$, and $F6$ with a specific $\alpha$ (e.g., $\alpha=0.3$) could result in relatively small $RD$s and $V\!AR$s on the whole.
    The reason is simple. Given a path, the expectation of its travel times or carbon emissions in $S$ scenarios generally have small difference for $2m+1$ scenario sets, especially in the scenario sets generated by SG, where the means of travel speeds on each link and each path are controlled to be equal to the means of the given distribution.

    \item When the objective function does not approximate the objective $F2$, both methods, but particularly the RS method, will be very sensitive to what happens to be the extreme values in a scenario set.
    As shown in the tables of $RD$ and $V\!AR$ results, on the whole, large $RD$s and $V\!AR$s are produced for $F1$ with large $\theta$, $F4$ with large $D$, $F5$ with large time interval $D-E$, and $F6$ with relatively small or large $\alpha$.
    Taking $F4$ with $D=9.00$ and $S=10$ as an example, the objective value of a path is equal to the mean of the tardiness $\max(D_p + T^s - D, 0)$ in 10 scenarios. In this case, the tardiness values in most scenarios are 0. These non-zero tardiness values, especially very large values from some extreme scenarios, have large effects on the value of $F4$.
    Then, even if we use SG to generate scenarios, the optimal objective values in $2m+1$ scenario sets would vary a lot, which results in large $RD$s and $V\!AR$s as shown in Table~\ref{tab:F4}.
    Even so, SG always generates smaller $RD$s and $V\!AR$s than RS does.

\end{enumerate}


\begin{table}
\caption{Comparison of $RD$ and $V\!AR$ as generated by RS and SG for O-D pair (65, 35) and different $\theta$ in $F1$.}
\resizebox{\linewidth}{0.32\linewidth}{	
{\begin{tabular}{cccccc} \toprule																			
     &\multirow{2}{*}{$\theta$}     & $S=10$ & $S=15$ & $S=20$ & $S=25$  \\												
     &          & RS/SG  & RS/SG  & RS/SG  & RS/SG \\ \midrule											
 \multirow{11}{*}{$RD$(\%)}	&	0&	(3.42, 5.60, 7.65)/0.44&	(2.6, 4.01, 6.67)/0.12&	(2.35, 3.12, 3.79)/0.23&	(1.60, 3.21, 5.84)/0.18	\\
&	0.2&	(3.62, 6.28, 9.25)/0.37&	(2.74, 4.56, 6.50)/0.23&	(2.54, 3.24, 3.92)/0.28&	(1.77, 2.52, 4.06)/0.30	\\
&	0.4&	(3.99, 7.18, 10.55)/0.35&	(2.68, 4.62, 6.47)/0.33&	(2.94, 3.60, 5.06)/0.33&	(2.13, 3.14, 4.58)/0.42	\\
&	0.6&	(5.49, 7.49, 9.96)/0.37&	(3.16, 4.62, 5.97)/0.44&	(2.55, 3.61, 4.98)/0.38&	(1.82, 3.38, 6.20)/0.54	\\
&	0.8&	(6.22, 7.64, 10.64)/0.49&	(3.93, 5.36, 12.21)/0.54&	(2.60, 4.55, 6.85)/0.45&	(2.52, 3.61, 6.60)/0.68	\\
&	1.0&	(4.52, 7.70, 11.64)/0.60&	(2.51, 6.00, 11.11)/0.64&	(2.44, 4.30, 5.75)/0.52&	(2.26, 3.56, 5.49)/0.81	\\
&	1.2&	(6.76, 8.96, 14.13)/0.70&	(4.35, 6.35, 9.45)/0.74&	(2.56, 4.56, 6.61)/0.59&	(2.40, 3.70, 4.79)/0.94	\\
&	1.4&	(4.02, 9.60, 14.57)/0.81&	(4.45, 6.39, 9.18)/0.85&	(4.69, 6.01, 8.02)/0.66&	(2.74, 4.40, 6.09)/1.07	\\
&	1.6&	(6.00, 10.47, 15.21)/1.10&	(4.92, 7.38, 10.09)/1.12&	(4.23, 5.44, 7.42)/0.86&	(3.12, 4.75, 6.61)/1.19	\\
&	1.8&	(5.27, 11.20, 16.89)/1.21&	(3.77, 6.66, 13.21)/1.23&	(4.30, 7.75, 12.00)/0.95&	(3.56, 5.37, 7.10)/1.32	\\
&	2.0&	(5.94, 11.49, 17.39)/1.32&	(5.29, 8.80, 14.16)/1.34&	(5.54, 7.13, 8.19)/1.03&	(4.06, 5.05, 6.57)/1.44	\\\midrule
\multirow{11}{*}{$V\!AR$}&	0&	(1729, 3945, 8040)/16&	(822, 2263, 6260)/2&	(775, 1081, 1430)/6&	(299, 1412, 4441)/4	\\
&	0.2&	(1629, 4892, 9596)/14&	(758, 2744, 5883)/7&	(676, 1310, 2288)/9&	(350, 809, 1672)/10	\\
&	0.4&	(2045, 6289, 11614)/16&	(994, 3131, 5906)/15&	(1131, 1590, 3063)/13&	(688, 1279, 2294)/20	\\
&	0.6&	(4058, 7606, 11472)/22&	(1334, 2870, 4753)/26&	(912, 1781, 2561)/19&	(309, 1527, 4478)/36	\\
&	0.8&	(5022, 9199, 18491)/32&	(1741, 4276, 18287)/42&	(967, 3140, 7607)/27&	(745, 2382, 10465)/56	\\
&	1.0&	(2356, 8720, 19638)/46&	(953, 6065, 18310)/60&	(741, 2258, 3339)/35&	(601, 1908, 4068)/81	\\
&	1.2&	(4653, 10649, 29516)/63&	(2527, 5918, 12948)/82&	(924, 3416, 6010)/46&	(684, 1915, 3017)/110	\\
&	1.4&	(2607, 14013, 31953)/85&	(2695, 6355, 15095)/108&	(2842, 4787, 8139)/57&	(1172, 2996, 4385)/144	\\
&	1.6&	(4424, 19466, 39131)/200&	(3185, 8294, 19264)/161&	(2136, 4227, 7570)/82&	(1305, 3462, 6783)/183	\\
&	1.8&	(4140, 21958, 43436)/251&	(2318, 8249, 31277)/196&	(2785, 9215, 19187)/99&	(1566, 4692, 8950)/227	\\
&	2.0&	(5939, 23145, 46295)/309&	(4783, 13340, 34110)/235&	(5807, 7419, 10353)/118&	(2376, 4046, 6920)/275	\\\bottomrule			\end{tabular}}}	
\label{tab:F1}							
\end{table}

\begin{table}
\centering
\caption{$ORD$(\%) values as generated by RS and SG for O-D pair (65, 35) and different $\theta$ in $F1$.}					
{\begin{tabular}{ccccc} \toprule																			
\multirow{2}{*}{$\theta$}  & $S=10$ & $S=15$ & $S=20$ & $S=25$  \\												
 & RS/SG  & RS/SG  & RS/SG  & RS/SG  \\ \midrule											
0&	0.25/0&	0.07/0&	0.02/0&	0.07/0\\
0.2&	0.25/0&	0.17/0&	0.08/0&	0.06/0\\
0.4&	0.39/0&	0.23/0&	0.09/0&	0.04/0\\
0.6&	0.42/0&	0.13/0&	0.11/0&	0.15/0\\
0.8&	0.51/0&	0.31/0&	0.27/0&	0.14/0\\
1.0&	0.72/0&	0.42/0&	0.25/0&	0.13/0\\
1.2&	0.69/0&	0.36/0&	0.28/0&	0.15/0\\
1.4&	0.54/0&	0.43/0&	0.40/0&	0.17/0\\
1.6&	0.88/0.02&	0.52/0.02&	0.24/0.05&	0.26/0.05\\
1.8&	0.79/0.04&	0.47/0.06&	0.58/0.08&	0.23/0.08\\
2.0&	0.78/0.04&	0.45/0.04&	0.46/0.05&	0.32/0.05\\ \bottomrule												
\end{tabular}}
\label{tab:ORD_F1}									
\end{table}


\begin{table}
\caption{Comparison of $RD$ and $V\!AR$ as generated by RS and SG for O-D pair (65, 35) and different $D$ in $F4$.}
\resizebox{\linewidth}{0.25\linewidth}{	
{\begin{tabular}{cccccc} \toprule																			
     &\multirow{2}{*}{$D$}     & $S=10$ & $S=15$ & $S=20$ & $S=25$  \\												
    &          & RS/SG  & RS/SG  & RS/SG  & RS/SG \\ \midrule											
 \multirow{8}{*}{$RD$(\%)}&	8.86&	(28.37, 38.32, 51.23)/4.03&	(19.58, 26.75, 36.75)/1.13&	(14.96, 22.26, 31.56)/2.05&	(15.40, 22.86, 29.92)/1.65	\\
&	8.88&	(22.38, 41.65, 57.92)/4.96&	(32.16, 39.06, 45.63)/1.40&	(16.96, 31.18, 45.24)/2.52&	(17.62, 25.48, 30.48)/2.03	\\
&	8.90&	(48.88, 54.16, 62.50)/4.54&	(16.53, 38.84, 60.79)/3.21&	(23.85, 34.95, 43.25)/4.44&	(16.08, 26.95, 37.95)/3.72	\\
&	8.92&	(58.72, 64.70, 72.90)/7.58&	(39.35, 47.69, 56.44)/3.71&	(24.84, 37.69, 50.07)/4.56&	(27.75, 40.52, 50.22)/5.78	\\
&	8.94&	(58.75, 72.57, 86.25)/9.37&	(38.90, 58.85, 74.62)/9.93&	(32.67, 48.32, 82.56)/6.03&	(32.7, 46.87, 66.27)/7.76	\\
&	8.96&	(66.33, 82.42, 100)/10.05&	(50.18, 68.31, 89.49)/14.68&	(49.43, 61.14, 77.11)/7.48&	(44.38, 54.41, 64.62)/12.67	\\
&	8.98&	(64.16, 92.08, 100)/29.57&	(76.93, 83.69, 100)/32.94&	(55.95, 77.96, 97.51)/24.26&	(50.11, 67.92, 86.36)/23.12	\\
&	9.00&	(94.06, 99.41, 100)/44.18&	(77.57, 90.42, 100)/47.88&	(68.28, 84.85, 99.61)/47.88&	(55.19, 75.86, 95.56)/42.20	\\\midrule
\multirow{8}{*}{$V\!AR$}&	8.86&	(1804, 4728, 9161)/16&	(808, 1526, 2966)/2&	(441, 997, 2237)/6&	(337, 959, 1446)/4	\\
&	8.88&	(767, 4505, 10135)/16&	(1128, 2376, 3986)/2&	(486, 1392, 2411)/6&	(332, 802, 1035)/4	\\
&	8.90&	(2764, 4176, 6071)/10&	(205, 1742, 4027)/6&	(439, 1165, 2023)/10&	(217, 595, 983)/6	\\
&	8.92&	(1630, 4762, 7981)/13&	(698, 1276, 2120)/5&	(374, 817, 1544)/6&	(407, 908, 1478)/9	\\
&	8.94&	(1033, 3771, 7302)/11&	(247, 1496, 3135)/16&	(283, 1278, 6537)/5&	(191, 598, 1188)/9	\\
&	8.96&	(784, 2252, 4859)/10&	(316, 1263, 3017)/16&	(282, 759, 1530)/4&	(250, 475, 824)/9	\\
&	8.98&	(290, 3034, 6625)/17&	(270, 704, 1284)/30&	(153, 558, 1874)/10&	(138, 377, 837)/12	\\
&	9.00&	(767, 1799, 4422)/17&	(79, 794, 2744)/17&	(203, 433, 1305)/16&	(63, 251, 1020)/16	\\\bottomrule								\end{tabular}}}	
\label{tab:F4}							
\end{table}

\begin{table}
\centering
\caption{$ORD$(\%) values as generated by RS and SG for O-D pair (65, 35) and different $D$ in $F4$.}					
{\begin{tabular}{ccccc} \toprule																			
  \multirow{2}{*}{$D$}  & $S=10$ & $S=15$ & $S=20$ & $S=25$  \\												
 & RS/SG  & RS/SG  & RS/SG  & RS/SG \\ \midrule											
8.86&	3.67/0&	1.24/0&	0.07/0&	0.10/0	\\
8.88&	2.77/0&	0.91/0&	0.17/0&	0.33/0	\\
8.90&	3.05/0&	1.14/0&	0.10/0&	0.20/0	\\
8.92&	5.26/0&	1.41/0&	0.58/0&	0.69/0	\\
8.94&	11.78/0&	3.52/0&	2.02/0&	0.74/0	\\
8.96&	13.08/0&	7.25/0&	4.94/0&	1.56/0	\\
8.98&	24.12/0&	8.40/0&	7.17/0&	5.37/0	\\
9.00&	23.60/1.04&	14.68/2.08&	11.19/4.15&	7.52/4.15	\\\bottomrule												
\end{tabular}}
\label{tab:ORD_F4}										
\end{table}


\begin{table}
\caption{Comparison of $RD$ and $V\!AR$ as generated by RS and SG for O-D pair (65, 35) and different time windows in $F5$.}
\resizebox{\linewidth}{0.27\linewidth}{	
{\begin{tabular}{cccccc} \toprule																			
    &Time     & $S=10$ & $S=15$ & $S=20$ & $S=25$  \\												
    &window   & RS/SG  & RS/SG  & RS/SG  & RS/SG \\ \midrule											
\multirow{9}{*}{$RD$(\%)}&	(8.90, 8.93)&	(48.17, 63.28, 82.01)/6.93&	(38.22, 49.31, 65.83)/6.93&	(31.79, 47.31, 59.09)/5.82&	(27.51, 40.09, 49.81)/8.06	\\
&	(8.87, 8.96)&	(69.55, 77.71, 95.98)/10.05&	(55.18, 71.28, 86.17)/14.68&	(52.44, 60.92, 68.19)/7.48&	(45.51, 57.32, 75.74)/12.67	\\
&	(8.84, 8.99)&	(85.21, 97.69, 100)/31.66&	(76.77, 86.73, 100)/38.07&	(67.15, 80.45, 96.81)/36.95&	(56.55, 74.08, 93.31)/32.26	\\
&	(8.84, 8.93)&	(50.23, 65.09, 92.09)/7.13&	(26.71, 44.01, 63.85)/5.18&	(30.49, 45.69, 52.91)/4.99&	(30.97, 39.66, 48.60)/6.48	\\
&	(8.81, 8.90)&	(42.61, 53.20, 63.03)/4.54&	(24.33, 44.11, 67.52)/3.21&	(31.15, 38.53, 43.61)/4.44&	(24.58, 29.63, 37.58)/3.72	\\
&	(8.78, 8.87)&	(22.64, 40.01, 50.80)/4.45&	(24.83, 32.00, 41.12)/1.25&	(21.74, 27.33, 30.67)/2.26&	(15.52, 24.02, 30.30)/1.82	\\
&	(8.9, 8.99)&	(84.19, 94.02, 100)/33.05&	(69.94, 83.11, 97.86)/40.25&	(70.17, 82.82, 95.02)/39.22&	(62.64, 70.51, 83.02)/34.06	\\
&	(8.93, 9.02)&	(88.43, 96.38, 100)/57.60&	(80.81, 87.27, 95.68)/54.06&	(79.64, 84.16, 91.44)/49.03&	(57.54, 73.68, 86.1)/44.31	\\
&	(8.96, 9.05)&	(96.17, 99.59, 100)/69.02&	(84.34, 93.54, 98.87)/41.07&	(82.76, 92.66, 99.74)/40.95&	(70.27, 82.29, 98.70)/50.47	\\\midrule

\multirow{9}{*}{$V\!AR$}&	(8.90, 8.93)&	(823, 3035, 9092)/10&	(589, 1091, 2089)/15&	(333, 1005, 1493)/8&	(285, 709, 1124)/11	\\
&	(8.87, 8.96)&	(715, 1546, 3218)/10&	(435, 1169, 1987)/16&	(452, 918, 3604)/4&	(220, 464, 1045)/9	\\
&	(8.84, 8.99)&	(739, 1545, 3607)/14&	(346, 566, 1151)/27&	(186, 667, 3046)/20&	(136, 275, 477)/15	\\
&	(8.84, 8.93)&	(880, 2371, 4783)/9&	(218, 1302, 5976)/7&	(407, 1037, 2809)/4&	(275, 535, 888)/8	\\
&	(8.81, 8.90)&	(2678, 4448, 7732)/10&	(590, 1806, 3897)/6&	(703, 1369, 2879)/10&	(371, 770, 1371)/6	\\
&	(8.78, 8.87)&	(764, 3554, 6803)/16&	(777, 1832, 2694)/2&	(611, 1188, 1634)/6&	(376, 858, 1334)/4	\\
&	(8.90, 8.99)&	(303, 1458, 2814)/20&	(209, 696, 1971)/32&	(135, 478, 804)/21&	(171, 338, 1041)/18	\\
&	(8.93, 9.02)&	(652, 1475, 2536)/42&	(357, 752, 1424)/33&	(195, 631, 1070)/28&	(73, 355, 780)/21	\\
&	(8.96, 9.05)&	(411, 1402, 4678)/57&	(434, 751, 1623)/29&	(147, 426, 1053)/25&	(177, 296, 845)/48	\\
\bottomrule												
\end{tabular}}}	
\label{tab:F5}								
\end{table}

\begin{table}
\centering
\caption{$ORD$(\%) values as generated by RS and SG for O-D pair (65, 35) and different time windows in $F5$.}
{\begin{tabular}{ccccc} \toprule																			
     Time    & $S=10$ & $S=15$ & $S=20$ & $S=25$  \\												
  window & RS/SG  & RS/SG  & RS/SG  & RS/SG  \\ \midrule											
(8.90, 8.93)&	3.95/0&	3.31/0&	1.57/0&	1.87/0	\\
(8.87, 8.96)&	6.26/0&	6.28/0&	3.97/0&	1.99/0	\\
(8.84, 8.99)&	19.19/1.58&	8.85/1.58&	9.55/1.58&	7.40/1.58	\\
(8.84, 8.93)&	4.22/0&	2.04/0&	1.41/0&	1.37/0	\\
(8.81, 8.90)&	4.37/0&	1.09/0&	0.50/0&	0.10/0	\\
(8.78, 8.87)&	1.59/0&	1.26/0&	0.52/0&	0.15/0	\\
(8.90, 8.99)&	18.33/2.94&	9.96/1.96&	7.58/1.96&	8.22/3.92	\\
(8.93, 9.02)&	31.92/8.62&	22.29/2.87&	26.33/2.87&	16.5/4.91	\\
(8.96, 9.05)&	19.03/4.01&	20.48/14.24&	16.29/13.17&	13.07/11.69	\\\bottomrule												
\end{tabular}}
\label{tab:ORD_F5}										
\end{table}	


\begin{table}
\caption{Comparison of $RD$ and $V\!AR$ as generated by RS and SG for O-D pair (65, 35) and different $\alpha$ in $F6$.}
\resizebox{\linewidth}{0.27\linewidth}{	
{\begin{tabular}{cccccc} \toprule																			
     &\multirow{2}{*}{$\alpha$}     & $S=10$ & $S=15$ & $S=20$ & $S=25$  \\												
     &          & RS/SG  & RS/SG  & RS/SG  & RS/SG \\ \midrule											
 \multirow{10}{*}{$RD$(\%)}	&	0.1&	(3.77, 7.93, 11.14)/4.17&	(4.84, 6.41, 8.74)/3.26&	(2.72, 4.98, 7.14)/3.01&	(3.28, 4.26, 5.31)/1.58	\\
&	0.2&	(5.67, 7.29, 10.89)/5.67&	(4.81, 5.81, 6.78)/3.25&	(4.03, 4.93, 6.84)/3.29&	(3.66, 4.48, 5.94)/3.80	\\
&	0.3&	(4.79, 6.87, 8.65)/3.45&	(4.27, 5.82, 7.50)/3.80&	(3.45, 4.64, 6.54)/3.80&	(2.65, 3.90, 5.29)/2.14	\\
&	0.4&	(5.52, 7.14, 10.35)/5.43&	(3.70, 5.55, 8.00)/4.29&	(3.31, 4.41, 5.33)/3.80&	(3.33, 4.21, 5.68)/2.09	\\
&	0.5&	(5.15, 8.28, 10.41)/5.30&	(1.89, 5.55, 10.07)/4.62&	(3.31, 5.39, 7.55)/1.10&	(2.11, 3.56, 5.29)/2.44	\\
&	0.6&	(5.54, 8.74, 11.86)/2.48&	(3.61, 5.64, 7.96)/2.47&	(2.91, 3.65, 5.12)/2.09&	(2.46, 3.68, 5.47)/2.22	\\
&	0.7&	(5.09, 8.85, 14.22)/3.61&	(5.32, 6.49, 8.56)/3.14&	(2.28, 4.58, 7.30)/2.48&	(2.07, 4.23, 6.36)/2.02	\\
&	0.8&	(7.66, 10.51, 14.45)/3.44&	(5.28, 7.06, 9.05)/2.40&	(3.91, 5.95, 7.73)/2.22&	(3.27, 5.01, 7.44)/2.62	\\
&	0.9&	(7.09, 13.25, 18.62)/4.67&	(6.33, 8.88, 14.13)/4.88&	(6.02, 7.31, 11.62)/3.87&	(4.66, 6.35, 8.43)/4.33	\\
&	1.0&	(9.52, 14.51, 18.62)/3.13&	(3.71, 9.96, 17.08)/2.65&	(8.67, 10.99, 12.86)/4.70&	(7.13, 11.07, 15.16)/4.59	\\\midrule
\multirow{10}{*}{$V\!AR$}&	0.1&	(2528, 6938, 14311)/1500&	(2428, 4616, 8732)/1030&	(766, 2801, 5351)/946&	(1146, 2164, 3939)/293	\\
&	0.2&	(3785, 6767, 11297)/4967&	(2912, 3902, 6002)/989&	(1750, 3169, 8657)/1017&	(1211, 2380, 3966)/1188	\\
&	0.3&	(2636, 5744, 12385)/1511&	(1719, 3871, 6902)/1889&	(1359, 2879, 3748)/1968&	(1115, 1955, 3241)/377	\\
&	0.4&	(3757, 5915, 9535)/3200&	(1044, 3571, 6408)/1803&	(1154, 2659, 4464)/1142&	(1066, 1999, 4406)/344	\\
&	0.5&	(2381, 7998, 12812)/3475&	(315, 3962, 9092)/2071&	(1324, 3522, 7207)/128&	(582, 1780, 4640)/802	\\
&	0.6&	(5763, 9744, 16466)/797&	(1435, 4511, 7469)/766&	(855, 1653, 3984)/573&	(736, 1663, 3199)/541	\\
&	0.7&	(3171, 10853, 27934)/1548&	(3073, 5213, 8311)/1049&	(668, 3173, 6593)/646&	(469, 2473, 4362)/613	\\
&	0.8&	(10611, 18373, 38065)/1838&	(4382, 7845, 12156)/818&	(2303, 5802, 9212)/713&	(1925, 4068, 8083)/917	\\
&	0.9&	(6558, 31329, 55924)/3430&	(5172, 15713, 41423)/4022&	(3622, 8377, 20615)/1562&	(2438, 6545, 16439)/2135	\\
&	1.0&	(20687, 43022, 61234)/1579&	(3652, 17016, 39717)/900&	(16007, 26113, 38528)/2910&	(11292, 25954, 44208)/2179	\\\bottomrule		\end{tabular}}}	
\label{tab:F6}									
\end{table}

\begin{table}
\centering
\caption{$ORD$(\%) values as generated by RS and SG for O-D pair (65, 35) and different $\alpha$ in $F6$.}
{\begin{tabular}{ccccc} \toprule																			
 \multirow{2}{*}{$\alpha$}           & $S=10$ & $S=15$ & $S=20$ & $S=25$  \\												
  & RS/SG  & RS/SG  & RS/SG  & RS/SG  \\ \midrule											
0.1&	0.55/0.80&	0.41/0.80&	0.39/0.53&	0.39/0.28	\\
0.2&	0.50/0.64&	0.53/0.31&	0.45/0.17&	0.37/0.17	\\
0.3&	0.32/0.02&	0.24/0.21&	0.19/0.21&	0.20/0.02	\\
0.4&	0.59/0.79&	0.36/0.46&	0.25/0.17&	0.45/0.05	\\
0.5&	0.50/0.44&	0.27/0.29&	0.36/0   &	0.17/0.15	\\
0.6&	0.68/0.18&	0.35/0.28&	0.21/0.18&	0.16/0	\\
0.7&	0.70/0.66&	0.27/0.21&	0.24/0.11&	0.24/0.05	\\
0.8&	0.79/0.31&	0.53/0.06&	0.31/0.04&	0.30/0.02	\\
0.9&	0.86/0.53&	0.53/0.38&	0.44/0.23&	0.37/0.25	\\
1.0&	3.91/2.45&	2.43/0.73&	2.12/1.22&	2.25/1.90	\\\bottomrule												
\end{tabular}}
\label{tab:ORD_F6}							
\end{table}

%

\end{document}